\documentclass[preprints,article,accept,moreauthors,pdftex]{Definitions/mdpi} 
\pdfoutput=1
\usepackage{amsmath}
\usepackage{amssymb}
\usepackage{graphicx}
\usepackage{tabularx}
\usepackage{bm}
\usepackage{tensor} 
\usepackage{physics}
\newcommand{\bea}{\begin{eqnarray}}
\newcommand{\eea}{\end{eqnarray}}
\newcommand{\be}{\begin{equation}}
\newcommand{\ee}{\end{equation}}

\firstpage{1} 
\pubvolume{1}
\issuenum{1}
\articlenumber{0}
\pubyear{2020}
\copyrightyear{2020}
\history{Received: date; Accepted: date; Published: date}

\Title{Towards black-hole singularity-resolution in the Lorentzian gravitational path integral}

\Author{Johanna N.~Borissova $^{1}$ \orcidB{0000-0001-7125-4372} and Astrid Eichhorn  $^{2}$\orcidA{0000-0003-4458-1495} }

\AuthorNames{Johanna N.~Borissova and Astrid Eichhorn}

\address{%
$^{1}$ \quad Rudolf Peierls Centre for Theoretical Physics, University of Oxford, Keble Road 1, Oxford OX1 3NP, UK; johanna.borissova@stcatz.ox.ac.uk\\
$^{2}$ \quad CP3-Origins, University of Southern Denmark, Campusvej 55, DK-5230 Odense M, Denmark; eichhorn@cp3.sdu.dk}

\abstract{Quantum Gravity is expected to resolve the singularities of classical General Relativity. Based on destructive interference of singular spacetime-configurations in the path integral, we find that higher-order curvature terms may allow to resolve black-hole singularities both in the spherically symmetric and axisymmetric case. In contrast, the Einstein action does not provide a dynamical mechanism for singularity-resolution through destructive interference of these configurations.}


\begin{document}

\section{Introduction}
Within General Relativity (GR), several solutions of the field equations develop curvature singularities. These signal the geodesic incompleteness of the corresponding spacetimes. The latter include the Schwarzschild, Kerr and Friedmann-Lema\^itre-Robertson-Walker (FLRW) metrics,  all of which are of relevance in astrophysics and cosmology. For instance, detections of gravitational waves from black-hole binaries \cite{Abbott:2016blz}, observations of stars orbiting the galactic center \cite{Ghez:2008ms} and the very first image of the shadow of a black hole \cite{Akiyama:2019cqa} show that astrophysical objects exist for which the spacetime metric is well-approximated by a Kerr metric, at least at sufficiently large geodesic distance to the object. Yet, that description must ultimately break down, as the curvature singularity of the Kerr metric cannot be physical. 
This singularity occurs at large values of the local curvature, where quantum-gravitational effects are expected to become important. Therefore, it is a key requirement of quantum gravity to provide a mechanism that reliably resolves such singularities. We assume, that even in the quantum gravitational regime, an effective metric description holds. Thus, the dynamical singularity-resolution mechanism in quantum gravity must result in regular black-hole spacetimes that agree with the Kerr spacetime at low curvature scales and are therefore candidate spacetimes that capture the true nature of the observed compact objects. 

We  search for such a mechanism within the gravitational path integral. Schematically, it reads
\begin{equation}
Z= \int \mathcal{D}g_{\mu\nu}\, e^{i S[g_{\mu\nu}]},\label{eq:PIgeneral}
\end{equation}
where all possible metric configurations (modulo diffeomorphisms) are being summed over.  Thus, the singular spacetime metrics that constitute solutions of the field equations in GR are included in the path integral. In addition, regular black-hole spacetimes, which agree (approximately) with the Schwarzschild or Kerr solution outside the black-hole horizon but do not harbor a curvature singularity, are also included. We ask whether there can be a dynamics that forces singular black-hole spacetimes to interfere destructively, suppressing the contribution of such spacetimes to the Lorentzian path integral, while regular black-hole solutions provide a finite contribution. The mechanism at work here, see \cite{Lehners:2019ibe}, is that any configuration $g_{\mu\nu}$ on which the action $S[g_{\mu\nu}]$ exhibits a divergence, is excluded from the path-integral through destructive interference. The divergence of the action signifies a rapidly oscillating quantum-mechanical phase factor for singular spacetimes. Thus, the singular spacetime and "neighboring" configurations (using an appropriate notion of distance on the configuration space) interfere destructively. \\

In this context it is important to highlight that in the path-integral expression \eqref{eq:PIgeneral}, $S$ need not be the Einstein-Hilbert action. Indeed, little is known about the form of the microscopic action $S$ in the path integral, as observations actually constrain the effective action $\Gamma$ that is obtained once the microscopic degrees of freedom are integrated over.
These observational constraints restrict the leading-order terms in a curvature/derivative expansion of $\Gamma$. For instance, the cosmological constant and the Newton coupling have been determined with good accuracy, while, at the quadratic order in the curvature, the observational constraints on the couplings are not very strong. The couplings are restricted to $\leq \order{10^{-61}}$ \cite{Calmet:2008tn}, using sub-millimeter tests of Newton's law \cite{Hoyle:2004cw}. Observational constraints on curvature invariants of cubic or higher order are not available yet.\\
 Since these \emph{observational} constraints all apply to the effective action, distinct, \emph{theoretical} constraints are required to fix the form of the microscopic action $S$. In fact, demanding a dynamical mechanism for singularity resolution through destructive interference can serve as one such theoretical constraint, as we will show in this paper. Indeed, requiring a divergence of the action on singular spacetimes divides actions into viable candidates and ruled-out dynamics, since a curvature singularity of a given spacetime is not necessarily reflected in a divergence of the action. This happens if the action is built out of only those curvature invariants that remain finite despite a divergence in some other curvature invariants. As a simple example, for the Schwarzschild black hole, the curvature invariant $R$ vanishes everywhere, while the invariant $R_{\mu\nu\kappa\lambda}R^{\mu\nu\kappa\lambda}$ diverges. Demanding a dynamics that diverges on singular black-hole spacetimes can therefore serve as a principle to distinguish different candidate dynamics for quantum gravity.
Accordingly, we will explore which form of the action is preferred in order to obtain a divergent action integral for different singular black-hole solutions. For cosmology, a similar investigation has been performed previously in \cite{Lehners:2019ibe} and has provided the intriguing result that curvature-squared terms could ensure a homogeneous and isotropic early universe. \\

A further important theoretical constraint is the predictivity of the theory, linked to its perturbative or non-perturbative renormalizability. 
Indeed, demanding control over ultraviolet (UV) divergences disfavors the Einstein-Hilbert action
\begin{equation}
S_{\rm EH}[g]=  \frac{1}{16\pi G_{\rm N}} \int \dd[4]{x}\sqrt{-g} \qty(R - 2 \Lambda),
\end{equation}
as it is not asymptotically free and in fact not even perturbatively renormalizable \cite{tHooftVeltman1974,Goroff1986,VandeVen1991}. Therefore, it cannot be predictive when used beyond the effective field theory regime \cite{Donoghue1994,Donoghue:1995cz}, since it features infinitely many free parameters linked to the infinitely many counterterms required for its perturbative renormalization. Additionally, it has been debated whether or not this dynamics can give rise to a well-defined Hartle-Hawking wavefunction, which would provide a well-defined beginning of the universe \cite{Hartle:1983ai,Feldbrugge:2017fcc,Halliwell:2018ejl}.
Both points motivate to go beyond the Einstein-Hilbert action.
 Generalizing to a curvature-squared action of the form
\begin{equation}\label{eq: Curvature-Squared-Action}
S[g] =  \int \dd[4]{x} \sqrt{-g} \left(\frac{1}{16\pi G_{\rm N}}\qty(R- 2 \Lambda)+ a R^2 + b R_{\mu\nu\kappa\lambda}R^{\mu\nu\kappa\lambda}\right),
\end{equation}
results in asymptotic freedom \cite{Stelle1977,Fradkin1982} for an appropriate choice of signs of the couplings, see \cite{Salvio:2018crh} for a review. 
Yet, unless $b=0$, this theory appears non-unitarity, as there is a spin-2 ghost entering the perturbative graviton propagator around a flat background, \cite{Stelle1977}, see \cite{Anselmi:2018ibi,Donoghue:2019ecz} for newer developments on this question.\\
Finally, the asymptotic-safety scenario \cite{Weinberg1979,Reuter:1996cp} motivates the presence of higher powers of curvature invariants in the action, see, e.g., \cite{Lauscher:2002sq,Machado:2007ea,Codello:2008vh,Benedetti:2009rx,Dietz:2012ic,Benedetti:2012dx,Falls:2013bv,Ohta:2015fcu,Demmel:2015oqa,Christiansen:2017bsy,Gonzalez-Martin:2017gza,Falls:2017lst,deBrito:2018jxt,Falls:2020qhj}, also in its unimodular form \cite{Eichhorn:2015bna}. At present, the form of the fixed-point action is not completely known; the so-called Reuter fixed point and its relevant directions are only determined within truncations of the full space of couplings, see \cite{Percacci2017,Eichhorn:2018yfc,ReuterSaueressig2019} for reviews. The most important limitation of these results in the present context is their origin in a Euclidean path integral, see \cite{Manrique:2011jc} for a first step in Lorentzian signature and \cite{Bonanno:2020bil} for a discussion of limitations and open questions.
Additionally, those results are obtained within the functional Renormalization Group (RG) framework, see \cite{Dupuis:2020fhh} for a review, which does not provide direct access to the microscopic action $S$ \cite{Manrique:2008zw,Manrique:2009tj,Morris:2015oca}. This motivates us to explore whether a requirement different from the fixed-point requirement can provide useful constraints on the form of the microscopic action. 

In summary, we explore whether in addition to the requirement of predictivity and (non-) perturbative renormalizability, a candidate mechanism for dynamical singularity-resolution of curvature singularities in black-hole spacetimes could also favor the presence of higher curvature terms in the action.\\

 This paper is structured as follows: In Sec.~\ref{subsec: BH metrics and scalars} we first motivate our choice of curvature invariants. We provide the line elements and values of the scalar invariants for various singular and non-singular black-hole spacetimes. In addition, relations between the curvature invariants that are particular for a given spacetime are highlighted. Sec.~\ref{subsec: Searching for SR dynamics} establishes our general ansatz for the  gravitational action and introduces functions for the spherically symmetric and axisymmetric spacetimes that are used for the subsequent analyses. Following the examination of minimal Einstein-Hilbert and curvature-squared dynamics for both singular and regular spacetimes, we turn to investigate beyond-four-derivative terms in the action. We conclude and provide a short outlook in Sec.~\ref{sec:conclusions}.

\section{Singular and regular black-hole spacetimes and their curvature invariants}\label{subsec: BH metrics and scalars}
We focus on two classes of black-hole metrics, namely those harboring a spacetime singularity, encoded in a divergent curvature invariant like the Kretschmann scalar, as well as regular ones where the curvature is finite everywhere. For our purposes, it is not relevant whether these black-hole spacetimes constitute solutions to the classical gravitational equations of motion, the Einstein equations, or to modified classical gravitational equations of motion. The gravitational path integral includes all configurations, irrespective of whether or not they solve Einstein's equations. In particular, we demand of a well-defined theory of quantum gravity, that singular black-hole spacetimes interfere destructively. On the other hand, we expect that with the appropriate boundary conditions, a regular black-hole spacetime can emerge from the Lorentzian gravitational path integral as the expectation value for the spacetime geometry -- in accordance with the observation of extremely compact objects in astrophysics. Therefore, we demand that i) singular black-hole spacetimes come with a divergent action, resulting in their destructive interference in the Lorentzian gravitational path integral and ii) regular black-hole spacetimes come with a finite action, allowing their contribution to the gravitational path integral.

We focus on a set of curvature invariants $K^{(i)}$ up to mass dimension eight, which is complete in a sense to be detailed below up to mass dimension six:
\bea
K^{(1)}&=&R,\nonumber\\  
K^{(2)}&=&R_{\mu\nu}R^{\mu\nu},\nonumber\\  
K^{(3)}&=&R_{\mu\nu\kappa \lambda}R^{\mu\nu\kappa\lambda},\nonumber\\ 
K^{(4)} &=& R\indices{^{\mu\nu\kappa\lambda}}R\indices{_\mu^\tau_\kappa^\omega}R\indices{_{\nu\tau\lambda\omega}},\nonumber \\
K^{(5)}&=& R^{\mu\nu\kappa\lambda}R_{\kappa\lambda\rho\sigma}R^{\rho\sigma}_{\,\,\,\,\,\mu\nu},\nonumber\\
K^{(6)} &=& R\indices{^{\mu\nu\kappa\lambda ; \tau}}R\indices{_{\mu\nu\kappa\lambda ; \tau}},\nonumber\\
K^{(7)}&=& R_{\mu\nu\kappa\lambda}R^{\kappa\lambda\rho\sigma}R_{\rho\sigma\alpha \beta}R^{\alpha\beta \mu\nu},\nonumber\\
K^{(8)}&=& (K^{(3)})^2. \label{eq: Scalars}
\eea
As has been shown in \cite{Narlikar}, a complete basis of non-derivative invariants can be constructed without including the dual Riemann tensor. Thus, we include invariants built solely out of the Riemann tensor and the metric, for which there are 14 independent ones in the most general case \footnote{One can see this by considering that the Riemann tensor has 20 independent components in four dimensions, and the metric has 10, but 16 of the 30 independent functions can be removed by a coordinate transformation, see, e.g., \cite{Overduin:2020aiq} for a discussion. Including invariants which contain the covariant derivative, additional invariants are present at each order in an expansion in derivatives and curvature.}, but only four independent non-vanishing ones for the case of an axisymmetric spacetime which is a vacuum solution to the Einstein equations \cite{Overduin:2020aiq}. We neglect topological invariants like the Gauss-Bonnet term and the Hirzebruch signature. Further, we neglect invariants which contain either $R$ or $R_{\mu\nu}$. It will become clear in our analysis that their inclusion would not alter our main result. 

At sixth order in derivatives, a complete set  includes both local invariants, which are built purely out of the Riemann tensor and its contractions, as well as derivative invariants which are built out of the Riemann tensor and its derivatives. Following \cite{Fulling1992}, there are two such local invariants at order 6 which do not include neither the Ricci scalar nor tensor and one invariant including two covariant derivatives and two Riemann tensors. On the black-hole spacetimes which are solutions to the vacuum Einstein equations, it holds that $K^{(4)}= 2 K^{(5)}$. In the more general case, these two differ and we therefore include both. Let us also highlight that $K^{(5)}$ is the Goroff-Sagnotti counterterm from two-loop perturbation theory of Einstein gravity \cite{Goroff1986}. 

At order four in the curvature, we only include two invariants in order to demonstrate the salient features of this order in the curvature.

\subsection{Singular black-hole spacetimes}
In the following, we will focus on three paradigmatic examples of singular black-hole spacetimes, namely the Schwarzschild black hole, Kerr black hole, and Vaidya spacetime. We will briefly comment on the case of the additional solutions that appear in classical curvature-squared gravity \cite{Lu:2015cqa,Lu:2015psa} in Sec.~\ref{subsec:curvaturesquared}. In the following, we provide the explicit form of the various curvature invariants for these metrics.

The Schwarzschild spacetime describes a spherically symmetric, static and asymptotically flat black hole in the GR vacuum with line element 
\begin{equation}\label{eq: LineElementSchwarzschild}
\dd{s^2} = -f\qty(r)\dd{t^2} + f\qty(r)^{-1}\dd{r^2} + r^2(\dd{\theta^2} + \sin^2{\theta}\dd{\phi^2}),
\end{equation}
and lapse function
\begin{equation}
f(r) = 1-\frac{2G_{\rm N} M}{r}.
\end{equation}
Here $G_{\rm N}$ is the classical Newton coupling and $M$ the black-hole mass measured by a distant observer. The existence of a coordinate singularity at $r=2G_{\rm N} M$ in these coordinates is irrelevant for our analysis, as we focus on curvature invariants.

As the spacetime is a vacuum solution to the field equations, the Ricci scalar vanishes, $R = K^{(1)}=0$, as do contractions of the Ricci tensor, e.g., $K^{(2)}=R_{\mu\nu}R^{\mu\nu}=0$. The lowest-order non-vanishing curvature invariant in a curvature expansion is the Kretschmann scalar,
\begin{equation}
K^{(3)}
= \frac{48G_{\rm N}^2M^2}{r^6},
\end{equation}
and its divergence at $r=0$ indicates that there is a curvature singularity at the center.
The higher-order, local curvature invariants can all be expressed in terms of $K^{(3)}$ following a simple dimensional argument as a direct consequence of the high degree of symmetry of the Schwarzschild spacetime. For instance, except for the numerical prefactor, the relation between $K^{(4)}$ and $K^{(3)}$ follows from their mass dimensionality,
\be
K^{(4)} = \frac{1}{\sqrt{48}}\left(K^{(3)} \right)^{3/2}.
\ee
Similarly,
\be
K^{(5)} = \frac{1}{\sqrt{12}} \left(K^{(3)} \right)^{3/2},
\ee
and finally
\be
K^{(7)}= \frac{1}{4} \left(K^{(3)}\right)^2.
\ee
The situation is different for the derivative invariants. While they can still be expressed as a function of $K^{(3)}$, the black-hole mass $M$ provides an additional mass scale that enters $K^{(6)}$, such that
\bea
K^{(6)} &=& \frac{720 G_{\rm N}^2 M^2}{r^9}\left(r-2G_{\rm N}\,M \right)\\
& =&\frac{5\, 3^{2/3}}{2\, 2^{1/3}}\left(K^{(3)} \right)^{4/3}\,\left(G_{\rm N} M\right)^{-2/3} -\frac{5\sqrt{3}}{8} \left(K^{(3)} \right)^{3/2} .\nonumber
\eea
$K^{(6)}$ is actually the lowest-order horizon-detecting invariant for the Schwarzschild spacetime, $\left(K^{(6)}\vert_{r=2G_{\rm N} M}\right)=0$, see \cite{Abdelqader:2014vaa}. 
\\

\begin{figure}[!t]
\includegraphics[width=\linewidth]{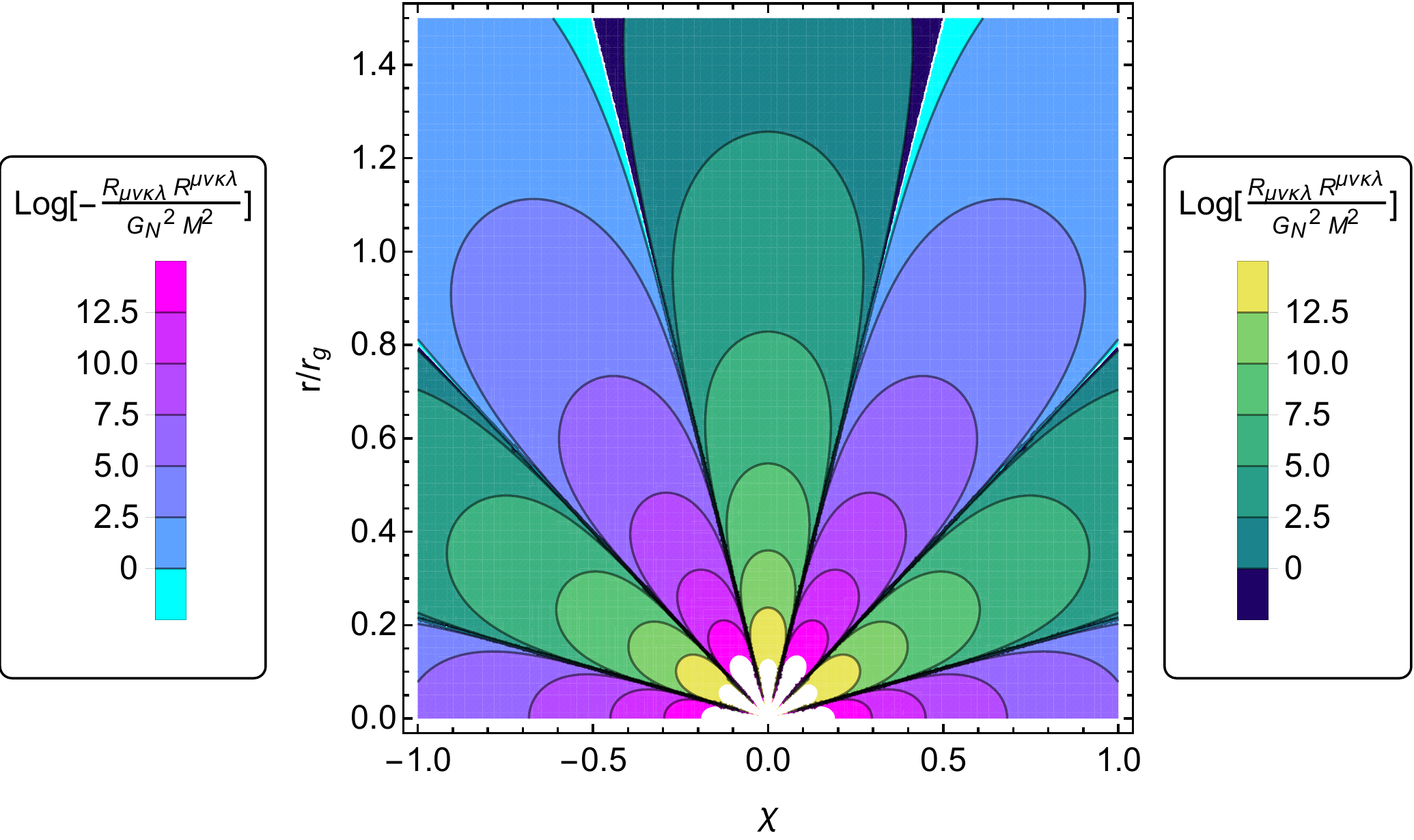}
\caption{We show the Kretschmann scalar as a function of $r$ and $\chi$ for the Kerr metric with spin parameter $a/r_g=0.8$. Positive and negative values alternate as a function of $\chi$.}
\label{fig: KretschmannKerr}
\end{figure}
The Kerr black hole is non-static and only axially symmetric due to its nonzero specific angular momentum $a=J/M$. 
It will be advantageous for the purpose of efficient evaluation of the curvature invariants \cite{Visser:2007fj} to work in rational polynomial coordinates, where $\cos(\theta) = \chi$, such that the line element is given by 
\bea
\dd{s^2} &=&  -\qty(1-\frac{2G_{\rm N}Mr}{\rho^2})\dd{t^2} - \frac{4G_{\rm N}M\,a\,r(1-\chi^2)}{\rho^2}\dd{t}\dd{\phi} + \frac{\rho^2}{\Delta}\dd{r^2} +\frac{\rho^2}{1-\chi^2}\dd{\chi^2}\nonumber\\
&{}&+ \left(1-\chi^2\right)(r^2+a^2+2G_{\rm N}M a^2 r \frac{(1-\chi^2)}{\rho^2})\dd{\phi^2},\label{eq:Kerrmetric}
\eea
where
\begin{equation}
\Delta =\Delta(r) = r^2 -2G_{\rm N}Mr +a^2,
\end{equation}
and 
\begin{equation}
\rho^2=\rho^2(r,\chi) = r^2 + a^2\chi^2.
\end{equation}
Just as in the Schwarzschild case, the first non-vanishing curvature invariant is
the Kretschmann scalar, given by 
\begin{equation}
K^{(3)}=  \frac{48 {G_{\rm N}}^2 M^2}{\qty(r^2+a^2\chi^2)^6}\left(r^6-15r^4a^2\chi^2 + 15r^2a^4\chi^4 - a^6\chi^6\right).
\end{equation}
The ring-singularity is located at $r=0$ in the equatorial plane at $\chi=0$, see Fig.~\ref{fig: KretschmannKerr}. In particular, there is a sequence of positive and negative curvature regions around the singularity. Taking the Schwarzschild limit $a\rightarrow 0$, the curvature becomes positive everywhere and diverges strongly in the entire $r=0$ surface.

The values of the non-vanishing dimension-six operators are given by
\bea
K^{(4)}&=&\frac{48G_{\rm N}^3M^3}{\qty(r^2+a^2\chi^2)^9}r\left(r^8-36r^6a^2\chi^2 +126r^4a^4\chi^4-84r^2a^6\chi^6+9a^8\chi^8\theta\right),\\
K^{(5)} &=& 2 K^{(4)},\\
K^{(6)}
&=  & \frac{720G_{\rm N }^2M^2}{\qty(r^2+a^2\chi^2)^9}\Big(r^8 -28r^6a^2\chi^2+70r^4a^4\chi^4 -28r^2a^6\chi^6+a^8\chi^8\Big) \qty(r\qty(r-2G_{\rm N}M)+a^2\chi^2).
\eea
They all exhibit the divergence characteristic for the Kerr spacetime, $K^{(i)}\rightarrow \infty$ for $3 \leq i \leq 6$ and $r \rightarrow 0$ at $\chi=0$. In contrast to the Schwarzschild case, $K^{(6)}$ is no longer horizon-detecting; instead, a particular combination of yet higher-order invariants plays this role \cite{Abdelqader:2014vaa}.

Finally, the dimension-six operator $K^{(7)}$ is given by
\be
K^{(7)}= \frac{576 G_{\rm N}^4M^4}{(r^2+a^2\chi^2)^{12}} \Bigl(r^{12}- 66 a^2 r^{10} \chi^2+495 a^4 r^8\chi^4-924 a^6 r^6 \chi^6+495 a^8 r^4 \chi^8 - 66 a^{10}r^2 \chi^{10}+a^{12}\chi^{12} \Bigr).
\ee

Further, we  consider the imploding Vaidya spacetime \cite{Vaidya1951, Vaidya1953, Vaidya1966, Kuroda1984} as an example for a singular non-vacuum spacetime. In Eddington-Finkelstein coordinates its line element can be expressed as
\begin{equation}
\dd{s^2} = -f\qty(r)\dd{v^2} + 2\dd{v}\dd{r} + r^2(\dd{\theta^2} + \sin^2{\theta}\dd{\phi^2}),
\end{equation}
where
\begin{equation}
f(r,v)= 1-\frac{2 G_{\rm N} M(v)}{r}.
\end{equation}
Here $M(v)$ is the mass function depending on the advanced time $v$. In particular, the spacetime is non-static and can be used to describe the formation of a black hole during the gravitational collapse of a radiative source \cite{Vaidya1966, Kuroda1984}. It represents an exact solution to the Einstein equations with energy-momentum tensor $T\indices{^\mu^\nu} = \rho k^\mu k^\nu$, where $\rho= (\partial_v M(v)) /4\pi r^2$ is the energy density. The fluid's four-velocity $k_\mu$ is a null-vector for radiation, thereby implying that the energy-momentum tensor has vanishing trace. Therefore, taking the trace of the field equations with vanishing cosmological constant shows that the Ricci scalar is zero,
\begin{equation}
K^{(1)}=0,
\end{equation} 
and also 
\be
K^{(2)}=0.
\ee
On the other hand, the Kretschmann scalar is given by
\begin{equation}
K^{(3)}
= \frac{48 {G_{\rm N}}^2 M^2(v)}{r^6}.
\end{equation}
Depending on the mass function, the spacetime may contain a singularity at the center.
The higher-order, local invariants are related to $K^{(3)}$ according to dimensional analysis,
\bea
K^{(4)}&=&48^{-1/2} \left(K^{(3)}\right)^{3/2},\\
K^{(5)}&=&12^{-1/2}\left(K^{(3)}\right)^{3/2},\\
K^{(7)}&=& \frac{1}{4}\left(K^{(3)} \right)^2.
\eea
Finally, the derivative invariant $K^{(6)}$ is the first invariant that depends on the derivative of the mass function,
\bea
K^{(6)}&=& - \frac{144 G_{\rm N} M(v)}{r^9} \Bigl(10G_{\rm N} M(v)^2 - 5 G_{\rm N} M(v) r + 2 r^2 M'(v) \Bigr).
\eea

\subsection{Regular black-hole spacetimes} \label{subsubsection: RegularBH-Spacetimes}
Regular black-hole spacetimes are currently being explored in terms of particular quantum-gravity approaches \cite{Bonanno2000,Ashtekar:2005qt,Modesto:2005zm,Bonanno:2006eu,Falls:2010he,Held2019,Platania:2019kyx,Faraoni:2020stz} as well as from a more agnostic point of view \cite{Bardeen1968,Dymnikova:1992ux,Hayward2005,Bambi:2013ufa,Frolov:2016pav}. This interest is in part triggered by novel observational opportunities, most importantly gravitational wave observations from binary black-hole mergers \cite{Abbott2016}, as well as imaging with the Event Horizon Telescope \cite{EHT2019,paper6}. Given that due to their curvature singularities the observed black holes \emph{cannot} be the Schwarzschild/Kerr solution from GR (even though observationally, deviations have not been detected yet and are in fact expected to be tiny outside the horizon based on a simple dimensional argument (see, however, \cite{Giddings:2019jwy} for alternatives), it becomes key to understand the true nature of the observed objects. Here, we are motivated by the possibility that they could be regular black-hole spacetimes. This requires that these provide a potentially finite contribution to the Lorentzian gravitational path integral and are not suppressed due to a divergent action.\\
As paradigmatic cases, we will focus on Hayward and Dymnikova spacetimes, as well as the finite-spin counterpart of a Hayward black hole. Within General Relativity, the stability of the inner horizons of such black-hole spacetimes has been discussed in \cite{Carballo-Rubio:2019nel,Bonanno:2020fgp}; we will ignore such dynamical questions here and only treat these spacetimes as some of many that have to be included in the path integral.

The Hayward black-hole spacetimes \cite{Hayward2005} are given by the line element \eqref{eq: LineElementSchwarzschild} with lapse function
\begin{equation}
f(r) = 1- 2G_{\rm N}M\qty(\frac{r^2}{r^3+2g^3}),
\end{equation}
where $g$ is a positive parameter. It plays the role of a transition scale between the Schwarzschild spacetime approximated at large radial distances and de Sitter space for small radii, $r\ll g$. At the center of the spacetime, curvature invariants are regular. The curvature invariants are given by
\bea
K^{(1)}
&=& \frac{24G_{\rm N}M}{\qty(r^3+2g^3)^3}\qty(-g^3r^3+4g^6),\\
K^{(2)}&=& \frac{288 G_{\rm N}^2 M^2 g^6}{(r^3+2g^3)^6}\left(5r^6-4r^3g^3+8g^6 \right),\\
 K^{(3)} &=&\frac{48 G_{\rm N}^2\, M^2}{(r^3+2g^3)^6} \left(r^{12}- 8 r^9 g^3+72 r^6\gamma^6-16 r^3 g^9+32 g^{12}\right),\\
 K^{(4)}&=&\frac{48 G_{\rm N}^3 M^3}{(r^3+2g^3)^7} \left(r^3-g^3 \right)\cdot \left(r^3-4g^3 \right)^3,
 \eea
 \bea
 K^{(5)}&=&\frac{96G_{\rm N}^3 M^3}{(r^3+2g^3)^9} \left(r^3-4g^3 \right)\cdot \Bigl(r^{15}-14g^3r^{12}+388 r^9g^6-424r^6g^9+32 r^3g^{12}-64 g^{15} \Bigr), \\
K^{(6)}&=& \frac{144 G_{\rm N}^2M^2 r^4}{(r^3+2g^3)^9} \left(r^2(r-2G_{\rm N} M)+2g^3 \right)\!\left(5 r^{12}-80r^9g^3+1200r^6g^6-1856r^3g^9+2432g^{12} \right).\nonumber\\
&{}&
\eea
Thus, the curvature
remains finite at the center and assumes the de Sitter value
\begin{equation}\label{eq: HaywardKretschmann}
\lim\limits_{r\to 0}
K^{(3)}
 = \frac{24{G_{\rm N}}^2M^2}{g^6}.
\end{equation}
As long as the de Sitter radius remains small enough, $g/r_g \leq \qty(4/3\sqrt{3})^{2/3}$, an outer event horizon continues to exist.
 It is worth noting that arguments from quantum gravity have been used to motivate Hayward black holes, see, e.g., \cite{Nicolini:2005vd,Rovelli:2014cta,Saueressig2015}.
For instance, the asymptotic-safety-inspired approach to resolve singularities based on a Renormalization Group improvement of classical line elements \cite{Bonanno2000, Falls:2010he,Litim:2013gga,Pawlowski:2018swz,Held2019} results in an effective Hayward spacetime. 

Next, we consider the static and spherically symmetric Dymnikova spacetime \cite{Dymnikova1992} with the line element \eqref{eq: LineElementSchwarzschild} and $r$-dependent mass function 
\begin{equation}
M(r)=M\qty(1-e^{-\frac{r^3}{2g^3}}).
\end{equation} 
Here, we only list two selected examples from the set of curvature invariants to highlight the difference to the Schwarzschild spacetime.
The Ricci and Kretschmann scalar are non-zero but finite,
\begin{equation}
K^{(1)}
= \frac{3G_{\rm N}M}{2g^6}e^{-\frac{r^2}{2g^3}}\qty(-3r^3+8g^3),
\end{equation}
\be
 K^{(3)}
 = \frac{3 G_{\rm N}^2\, M^2}{4 r^6g^{12}}
\left( 27 r^{12} - 48 \left( -2+e^{\frac{r^3}{2g^3}}\right)r^6g^6 - 64 \left(-1+e^{\frac{r^3}{2g^3}} \right)r^3g^9+64 \left(-1+e^{\frac{r^3}{2g^3}} \right)^2g^{12}
\right).
\ee
At the center, the curvature reduces to that of de Sitter space, just as for the Hayward spacetime \eqref{eq: HaywardKretschmann}, whereas it is asymptotically Schwarzschild at large $r$. The Dymnikova metric has also been motivated as a model for black holes in asymptotically safe gravity \cite{Platania:2019kyx}.

\begin{figure}[!t]
\centering
\includegraphics[width=0.6\linewidth]
{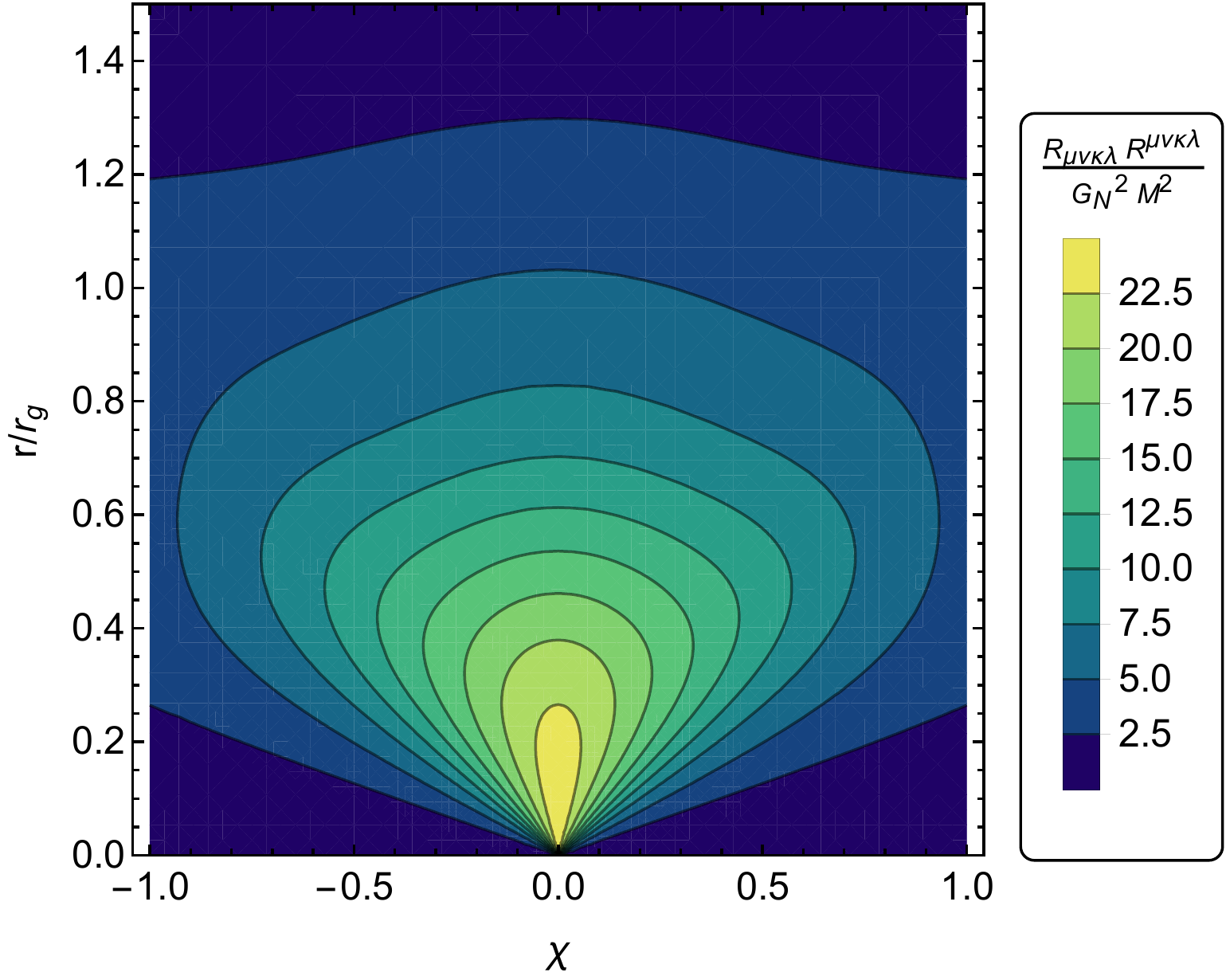}\\
\caption{We show the Kretschmann scalar as a function of $r$ and $\chi$ for the rotating Hayward metric in Eq.~\eqref{eq:Kerrmetric} with mass function given by \eqref{eq:MassHayward}  with spin parameter $a/r_g=0.5$ and Hayward parameter $g/r_g=1.0$.}
\label{fig: KretschmannRotatingHayward}
\end{figure}

\begin{figure}[!t]
\centering
\includegraphics[width=0.6\linewidth]{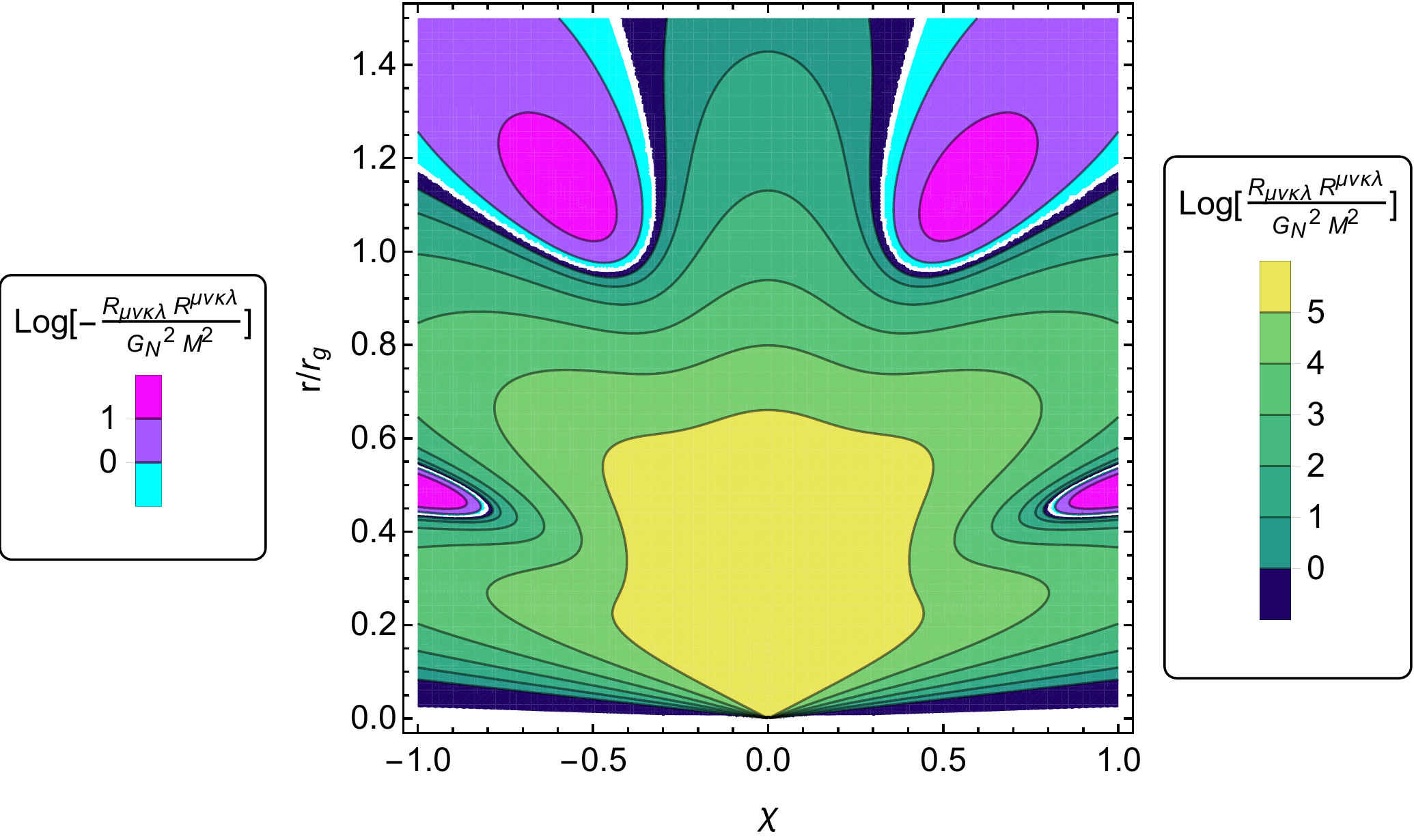}\\
\includegraphics[width=0.6\linewidth]{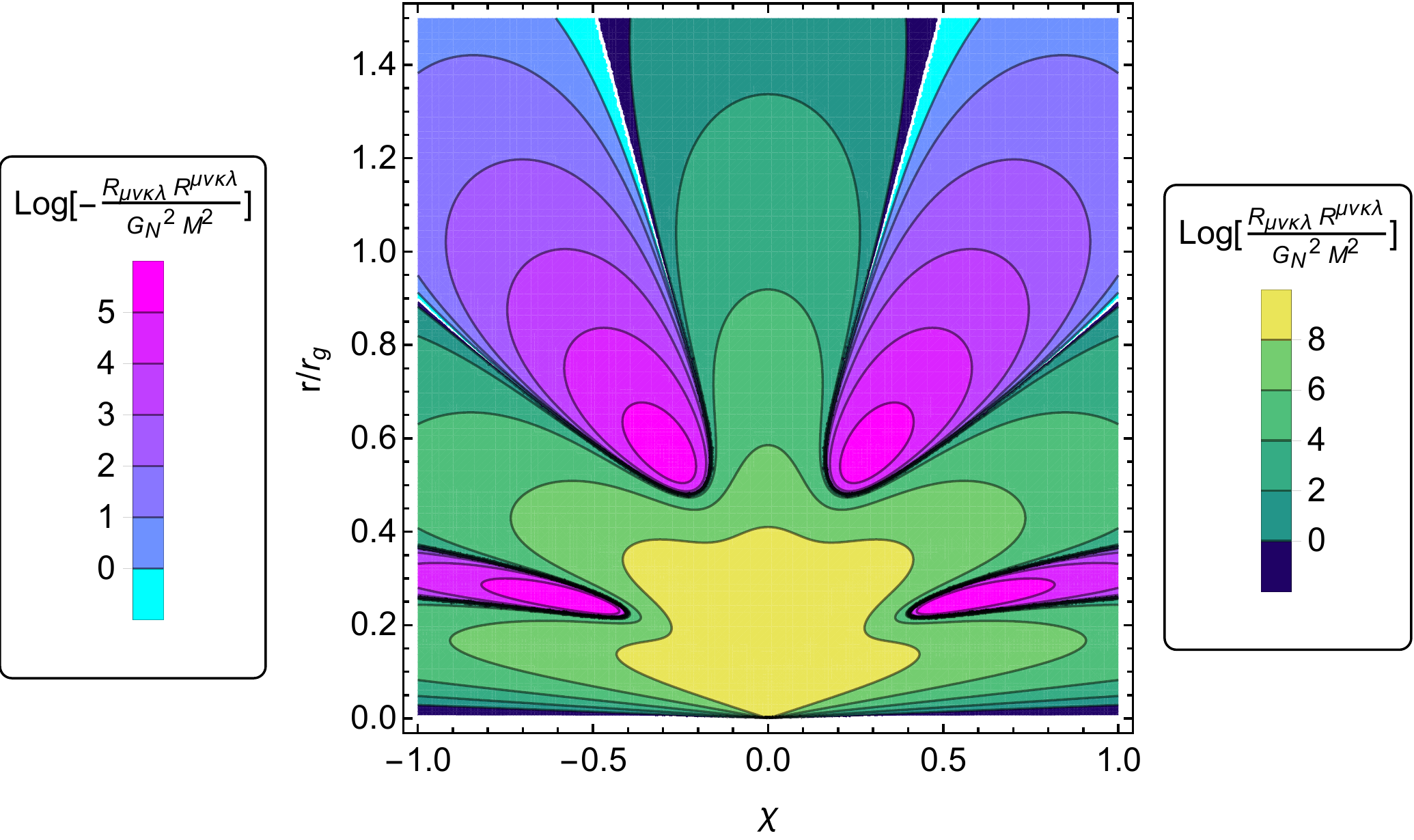}\\
\includegraphics[width=0.6\linewidth]{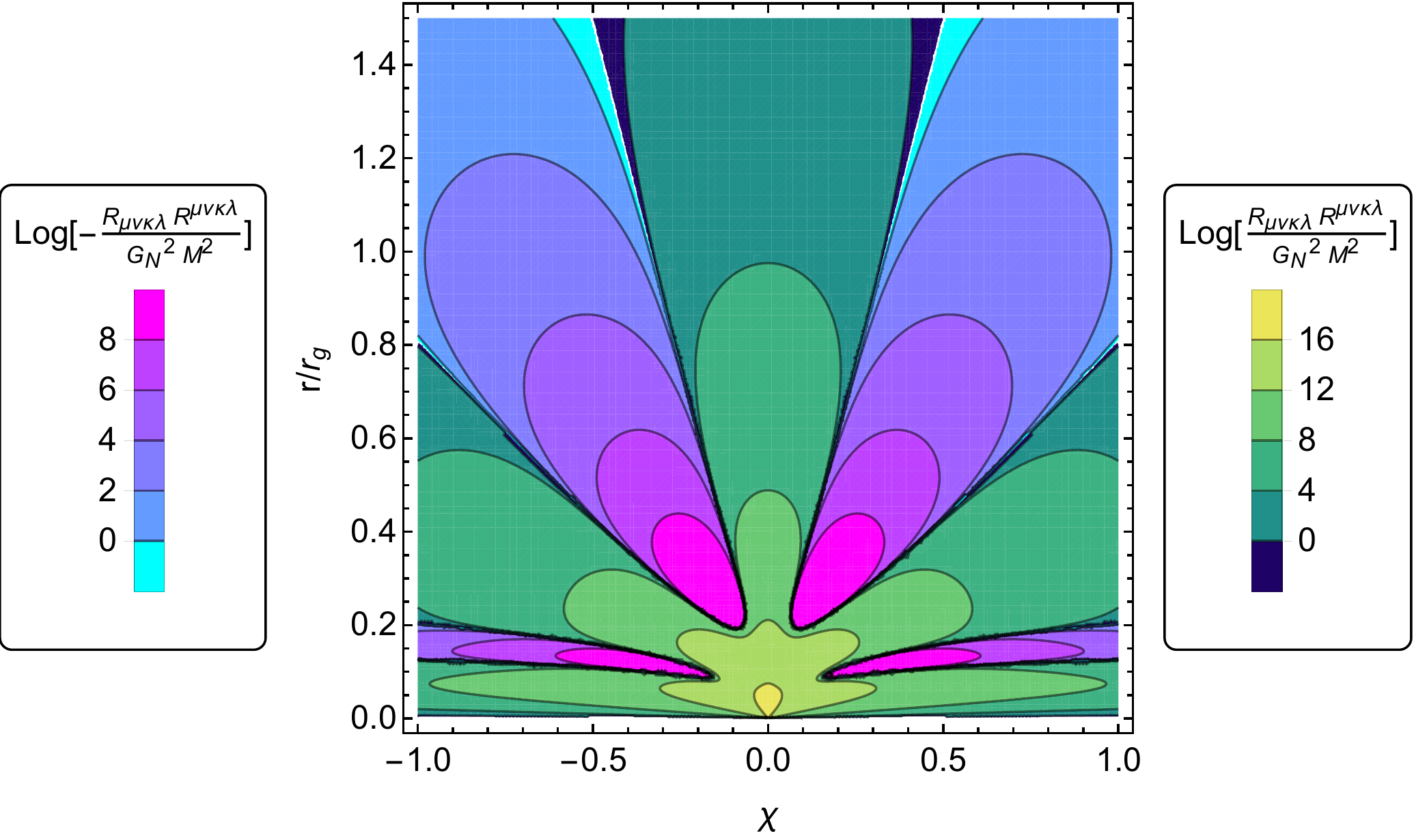}
\caption{We show the Kretschmann scalar as a function of $r$ and $\chi$ for the rotating Hayward metric in Eq.~\eqref{eq:Kerrmetric} with mass function given by \eqref{eq:MassHayward}  with spin parameter $a/r_g=0.5$ and Hayward parameter $g/r_g=0.5$ (upper panel), $g/r_g=0.25$ (central panel) and $g/r_g=0.1$ (lower panel). Decreasing values for the Hayward parameter lead to negative-curvature regions approaching the point $r=0, \chi=0$ from further outwards. Note the changing plotranges in the three panels.}
\label{fig: KretschmannRotatingHayward2}
\end{figure}

Our final example for regular black holes is the finite-spin counterpart of a Hayward black hole. It is obtained by the application of the Newman-Janis algorithm to the static Hayward spacetime, see \cite{Bambi2013, Newman1965,Drake:1998gf}. It results in a metric given by the Kerr-metric Eq.~\eqref{eq:Kerrmetric} with $r$-dependent mass function,
\begin{equation}
M(r)= M\qty(\frac{r^3}{r^3+2g^3}).\label{eq:MassHayward}
\end{equation}
The resulting spacetime is regular everywhere,
\begin{equation}
K^{(1)}
= \frac{24G_{\rm N}M\,g^3}{\qty(r^2+a^2\chi^2)\qty(r^3+2g^3)^3}r^2\qty(-r^3+4g^3),
\end{equation}
\bea
K^{(2)}&=& \frac{288 G_{\rm N}^2 M^2 r^4 g^6}{(r^3+2g^3)^6(2r^2+a^2\chi^2)^4} \cdot \Bigl( r^4(8g^6-4g^3 r^3+5 r^6) + 4r^2 \,a^2\, \chi^2(4g^6-5g^3r^3+r^6) \nonumber\\
&{}&+a^4\chi^4(r^3-4g^3)^2\Bigr),
\eea
and
\bea
K^{(3)}
&= & \frac{48{G_{\rm N}}^2M^2}{\qty(r^2+a^2\chi^2)^6\qty(r^3+2g^3)^6}\Biggl(32r^{12}\qty(r^3+2g^3)^4 -48r^{10}\qty(r^2+a^2\chi^2)\qty(r^3+2g^3)^3 \qty(r^3+ 4g^3)\nonumber\\
& {}&+ 12g^6\qty(r^3-4g^3)^2 r^4\qty(r^2+a^2\chi^2 )^4 + 2r^8\left(r^2 +a^2\chi^2\right)^2\qty(r^3+2g^3)^2  \qty(9r^6+76g^3r^3+212g^6) \nonumber\\
&{}& - r^6\qty(r^2+a^2\chi^2)^3\qty(r^3+2g^3)\big(r^9+6g^3r^6 + 72g^6r^3+416g^9\big)\Biggr).
\eea
We do not provide the explicit expressions for the higher-order invariants due to their lengthiness.

At the spacetime points where the Kerr metric exhibits a singularity, $(r,\chi) \rightarrow (0,0)$, the curvature remains finite. However, the values of some of the curvature invariants at $(r=0, \chi=0)$ depend on how this point is approached, see also \cite{Bambi2013},
\begin{equation}
\begin{split}
\lim\limits_{r\to 0}\lim\limits_{\chi\to0}
K^{(1)}
= {}& \frac{12 G_{\rm N} M}{g^3}, \\
\lim\limits_{\chi \to 0}\lim\limits_{r\to 0}
K^{(1)}
= {}& 0.
\end{split}
\end{equation}
A similar behavior can be observed for $K^{(2)}, K^{(3)}, K^{(4)}$ and $K^{(5)}$. On the other hand, $K^{(6)} \rightarrow 0$ as $r \rightarrow 0$, independent of whether one restricts to the equatorial plane first.

Fig.~\ref{fig: KretschmannRotatingHayward} shows the Kretschmann curvature for a finite choice of the rotation parameter $a$ and the Hayward parameter $g$. For smaller values of the spin, the contour areas of constant curvature extend along larger ranges of the angular coordinate. Thus, spherical symmetry is approximated.\\
 At finite values of the spin parameter, smaller values of the Hayward parameter lead to a strong increase of the the absolute value of the Kretschmann scalar along the radial and angular direction towards the critical point $(r,\chi)=(0,0)$. This is consistent with the Kerr singularity occurring for $g\rightarrow 0$. At the same time, regions of negative Kretschmann scalar move in towards $(r,\chi)=(0,0)$, cf.~Fig.~\ref{fig: KretschmannRotatingHayward2}. 

\section{Singularity-resolving dynamics}\label{subsec: Searching for SR dynamics}
In the following, we work with a gravitational action of the form
\begin{equation}\label{eq: ActionTerm}
S[g] = \sum_{i=1}^N g_i \int_V \dd[4]{x}\sqrt{-g} K^{(i)},
\end{equation}
where the $g_i$ denote the corresponding couplings.
We will start with the simple choice $S = S_{\rm EH}$, for which $N=1$, and then include progressively higher invariants, until we find that the spacetime integral results in a divergence. In other words, we will use the main idea from \cite{Lehners:2019ibe}
in a slightly different way: demanding that singular spacetimes exhibit a divergence allows us to differentiate between different candidates for gravitational actions.
\\ 
Here, we are not interested in divergences that arise from the infrared (IR), i.e., large distances, such as one would encounter when integrating the spacetime volume for the entire Minkowski spacetime. Instead, we will introduce an appropriate IR cutoff, indicated by the subscript $V$ in the spacetime integral in Eq.~\eqref{eq: ActionTerm}. In the presence of symmetries, an integration over a non-compact direction associated to a Killing vector trivially results in a (potentially IR-divergent) prefactor, but carries no information on dynamical singularity resolution. Therefore, we will leave out the corresponding integration.
For instance, we drop the time integration in the case of a static or stationary metric, since it can be split-off trivially from the integral over curvature invariants, and leads to an IR divergence unless a finite boundary is introduced. Any such constants and/or factors arising from these integrations are omitted in the following. We only carry out the integration over the coordinates the curvature invariant $K$ depends on.\\
For the spherically symmetric case, all curvature invariants depend on the radial distance to the origin only, i.e., in our choice of coordinates $K^{(i)}=K^{(i)}(r)$. On this basis we define the functions
\begin{equation}\label{eq: ActionFunctionSchwarzschildType}
{S_{\rm sph}}^{(i)}(r_{\rm UV}, r_{\rm IR}) = \int^{ r_{\rm IR}}_{r_{\rm UV}} \dd{r'}r'^2 K^{(i)}(r').
\end{equation}
The integral runs from a UV cutoff, denoted by $r_{\rm UV}$, to an IR cutoff, $r_{\rm IR}$. It will turn out that the limit $r_{\rm IR} \rightarrow \infty$ can be taken trivially, as all spacetimes we consider are asymptotically flat and therefore the upper integration boundary does not contribute.
We are interested in the limit $r_{\rm UV} \rightarrow 0$, where curvature singularities occur in classical black-hole spacetimes which may lead to divergences in the corresponding actions.

For the axially symmetric case one has to be more careful, since the singularity in the Kerr spacetime lies at $r\rightarrow 0$ and $\chi \rightarrow 0$, i.e., in the middle of the integration region for the $\chi$-integral. Accordingly, we define the action as
\bea 
S_{\rm ax}^{(i)}(r_{\rm UV}, r_{\rm IR}, \epsilon) 
&=& \int_{r_{\rm UV}}^{r_{\rm IR}}dr' \Biggl( \int_{-1}^{-\epsilon} d\chi' (r'^2 + a^2 \chi^2) K^{(i)}(r',\chi') + \int_{\epsilon}^1 d\chi' (r'^2 + a^2 \chi^2) K^{(i)}(r',\chi') \Biggr).
\label{eq: ActionFunctionKerrType}
\eea
The limits $r_{\rm UV} \rightarrow 0$ and $\epsilon \rightarrow 0$ have to be taken with some care, while the limit $r_{\rm IR} \rightarrow \infty$ is again trivial due to asymptotic flatness.

The values of the functions \eqref{eq: ActionFunctionSchwarzschildType} and \eqref{eq: ActionFunctionKerrType} in general carry non-zero mass dimension, since they only contain the salient part of the action integral and we do not account for the dimensionful coupling constants in front of the curvature invariants in the Lagrangian. The mass dimensions of our action functions depend on the curvature invariant $K^{(i)}$.
In the following, \eqref{eq: ActionFunctionSchwarzschildType} and \eqref{eq: ActionFunctionKerrType} are nevertheless referred to as the 'action functions'.

\subsection{Einstein-Hilbert dynamics}
We first focus on the Einstein-Hilbert action, i.e.~we set $N=1$ and consider only $K^{(1)}=R$ in the action \eqref{eq: ActionTerm}.

For a vacuum solution to the Einstein equations with vanishing cosmological constant, the Ricci scalar vanishes, thus, Ricci flatness applies to the Schwarzschild and the Kerr black hole. For the imploding Vaidya spacetime, the four-velocity of the ingoing radiation is a null-vector, such that the Ricci scalar also vanishes. Therefore an Einstein-Hilbert dynamics is ruled out as a viable choice to enable the destructive interference of all singular black-hole metrics in the path integral. \\
Instead, we are led to conclude that terms beyond the Ricci scalar should be added to the microscopic gravitational action, in order to provide a potential mechanism for destructive interference of singular spacetimes. Thus, there are two independent lines of arguments against the Einstein-Hilbert action as the microscopic action for gravity. First, it does not give rise to an asymptotically free or safe theory, instead requiring the introduction of a finite new-physics scale, as also indicated by its perturbative non-renormalizability \cite{tHooftVeltman1974,Goroff:1985sz,vandeVen:1991gw}. Second, it is insufficient to suppress singular cosmological spacetimes \cite{Lehners:2019ibe}, and as argued here, singular black-hole spacetimes.\\

\subsection{Curvature-squared dynamics}\label{subsec:curvaturesquared}
Accordingly, higher-derivative terms should be included into the action. In this subsection, we study the impact of curvature-squared invariants on the behavior of the action around the singularity. Due to the Gauss-Bonnet identity in four dimensions, the combination $\sqrt{-g}\qty(R_{\mu\nu\kappa\lambda}R^{\mu\nu\kappa\lambda}-4 R_{\mu\nu}R^{\mu\nu}+R^2)$ forms a total derivative and can be integrated to provide a topological invariant, which we do not consider in our action. Thus, we limit ourselves to adding $K^{(2)}$ and $K^{(3)}$ to the Einstein-Hilbert action.
Then the action \eqref{eq: ActionTerm} becomes
\bea\label{eq: CurvatureSquaredAction} 
S_{\rm quad}[g] \!=\!  \!\int \dd[4]{x} \!\sqrt{-g} \!\left(\frac{K^{(1)}}{16\pi G_{\rm N}}
+ g_2 K^{(2)} 
+ g_3 K^{(3)}
\right).
\eea
For black-hole metrics which are solutions to the Einstein equations, all but the last term in Eq.~\eqref{eq: CurvatureSquaredAction} vanish; thus we focus on the Kretschmann scalar, $K^{(3)}$.
For the Schwarzschild black hole, the action function \eqref{eq: ActionFunctionSchwarzschildType} diverges in the vicinity of $r=0$, since,
\begin{equation}
\lim\limits_{r_{\rm UV}\to 0}{S_{\rm sph}}^{(3)}(r_{\rm UV}, r_{\rm IR}\rightarrow \infty) = \lim\limits_{r_{\rm UV}\to 0}\frac{16{G_{\rm N}}^2M^2}{r_{\rm UV}^3}.\label{eq:S_Schw}
\end{equation} 
For the Vaidya metric, the integral over the advanced time only contributes by an overall prefactor to the spacetime integral. Therefore the dependence of the action on the radial coordinate is the same as for the Schwarzschild spacetime, resulting in a UV divergence and a corresponding suppression in the path-integral.\\

Let us also add that for the additional singular black-hole metrics that appear as the solution to the classical equations of motions in curvature-squared gravity, the Kretschmann scalar scales like $r^{-6}$ (for the so-called (1,-1)-family ) and like $r^{-8}$ (for the so-called (2,2) family) near the origin, respectively \cite{Lu:2015cqa,Lu:2015psa}, while the volume element scales like $r^2$. Accordingly, an action containing $K^{(3)}$ will feature a UV divergence on these solutions.\\

Next, we consider the Kerr metric. Let us first show the result of integrating $\sqrt{g} K^{(3)}$ in the equatorial plane, where
\bea
\int_{r_{\rm UV}}^{r_{\rm IR} \rightarrow \infty} \sqrt{-g}\, K^{(3)}\Big|_{\chi=0} = \frac{16 G_{\rm N}^2 M^2}{r_{\rm UV}^3},
\eea
which diverges as $r_{\rm UV} \rightarrow 0$. We expect to rediscover this divergence once we also integrate over $\chi$, where all angular directions except $\chi=0$ are expected to provide a non-singular contribution to the action function.
Due to the singularity in the equatorial plane, we have to split the integral over the Kretschmann scalar as in Eq.~\eqref{eq: ActionFunctionKerrType} and obtain:
\bea
\underset{r_{\rm UV} \rightarrow 0}{\rm lim} \,\underset{\epsilon \rightarrow 0}{\rm lim}S_{\rm ax}^{(3)} (r_{\rm UV}, \infty, \epsilon) &=&16\, G_{\rm N}^2\, M^2 \underset{r_{\rm UV} \rightarrow 0}{\rm lim} \,\underset{\epsilon \rightarrow 0}{\rm lim}\Biggl( \frac{r_{\rm UV}^3-r_{\rm UV} a^2}{(r_{\rm UV}^2+a^2)^3}- \epsilon\, r_{\rm UV}\frac{r_{\rm UV}^2- a^2\epsilon^2}{(r_{\rm UV}^2 + a^2 \epsilon^2)^3}\nonumber\\
&{}&+\frac{r_{\rm UV}^3 -r_{\rm UV} a^2}{(r_{\rm UV}^2 + a^2)^3} - \epsilon\, r_{\rm UV} \frac{r_{\rm UV}^2 - a^2 \epsilon^2}{(r_{\rm UV}^2 + a^2 \epsilon^2)^3}\Biggr),
\eea
where the contributions vanish in the limit $r_{\rm UV} \rightarrow 0$ except for parts of the contributions from the boundary that approaches the ring-singularity. The vanishing of the remaining contributions is due to the fact that the Kretschmann scalar changes sign between $\chi=-1$ and $\chi=1$ several times, cf.~Fig.~\ref{fig: KretschmannKerr}. In the above expressions, taking the naive limits $r_{\rm UV} \rightarrow 0$ at finite $\epsilon$ or $\epsilon \rightarrow 0$ at finite $r_{\rm UV}$ results in a vanishing result. To properly account for the effects of the ring singularity, the limits $\epsilon \rightarrow 0$ and $r_{\rm UV} \rightarrow 0$ should be taken concertely, i.e., we  define $r_{\rm UV} = \gamma\, \epsilon$ with some finite $\gamma$, such that taking $\epsilon \rightarrow 0$ allows us to approach the ring-singularity. As a result, we obtain
\bea
\underset{\epsilon \rightarrow 0}{\rm lim}\, S_{\rm ax}^{(3)}(\gamma \epsilon, \infty, \epsilon) &=& \underset{\epsilon \rightarrow 0}{\rm lim} \,32\, \gamma\, G_{\rm N}^2\, M^2 \frac{\gamma^2 - a^2}{(\gamma^2-a^2)^3}\frac{1}{\epsilon^2}.
\eea

Therefore, we conclude that the inclusion of the Kretschmann scalar in the Lorentzian action provides a mechanism to dynamically suppress curvature singularities. \\

Conversely, regular black-hole spacetimes should not be ruled out from the path integral.
In the following, we therefore verify the finiteness of the integral over the curvature of the regular black-hole spacetimes. This must be independent of the chosen boundary conditions. In particular, the action function for regular black holes must be regular at the point where the Schwarzschild and Kerr spacetimes are singular.\\
The action function \eqref{eq: ActionFunctionSchwarzschildType} with the Kretschmann scalar of the static Hayward black hole assumes the form
\begin{equation}
{S_{\rm sph}}^{(3)}(r_{\rm UV}) =  \frac{16{G_{\rm N}}^2 M^2}{5 \qty(r_{\rm UV}^3+2g^3)^5}\Big(5r_{\rm UV}^{12}+120g^6r_{\rm UV}^6 +100g^9r_{\rm UV}^3+72g^{12}\Big).
\end{equation}
It is evident that a non-zero Hayward parameter $g$ ensures a finite and non-zero value of the action function when evaluated over the entire $r$-range,
\begin{equation}
\lim\limits_{r_{\rm UV}\to 0}{S_{\rm sph}}^{(3)}(r_{\rm UV}) = \frac{36 {G_{\rm N}}^2 M^2 }{5 g^3}.
\end{equation}
For the Dymnikova-spacetime, we obtain a similar result,
\begin{equation}
\lim\limits_{r_{\rm UV}\to 0}{S_{\rm sph}}^{(3)}(r_{\rm UV}) = \frac{27 {G_{\rm N}}^2 M^2 }{2g^3}.
\end{equation} 

\begin{figure}
\includegraphics[width=\linewidth]{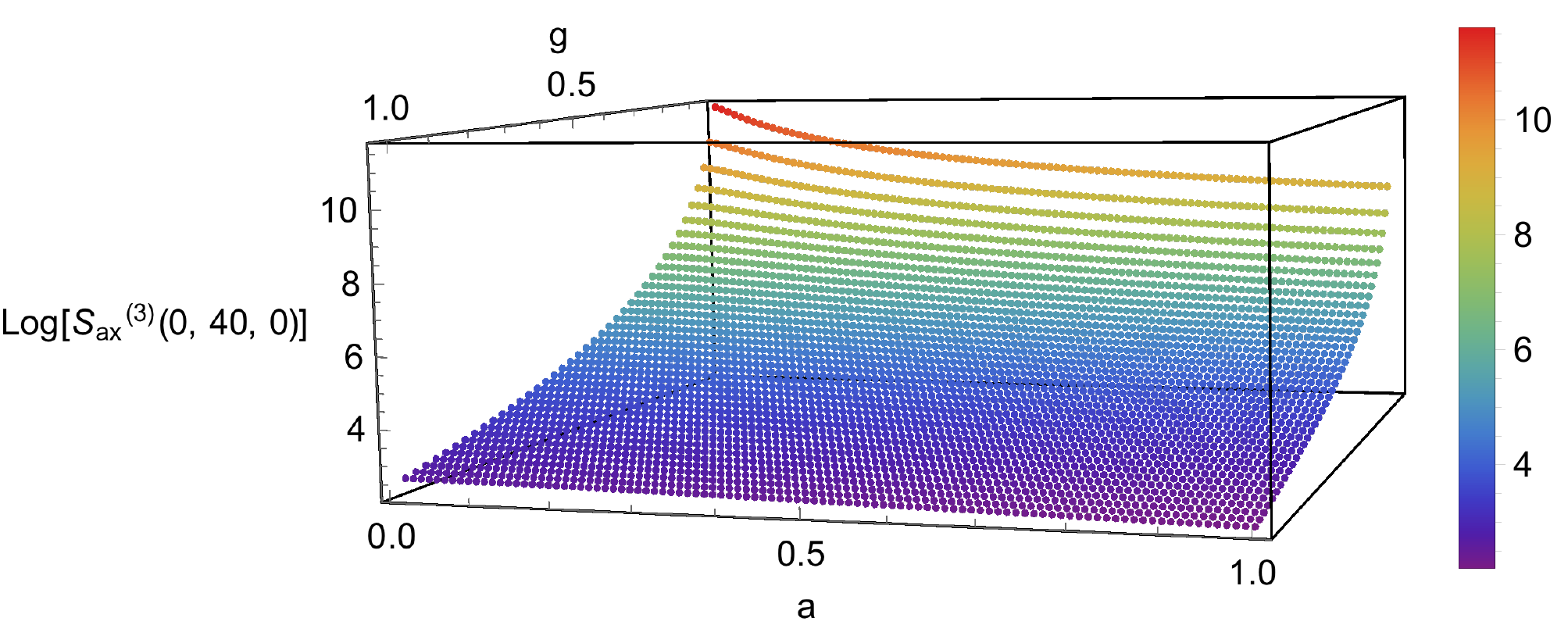}
\caption{\label{fig:rotHaywardint} We show the value of the logarithm of the numerical integral for the action function (in appropriate units) as a function of Hayward parameter $g$ and specific angular momentum $a$. The expected strong increase of the action function for $a \rightarrow 0$ and $g \rightarrow 0$ is clearly visible.
Note that the smallest Hayward parameter and smallest spin parameter we consider are $g=0.05$ and $a=0.01$, respectively, such that the action function is always finite.}
\end{figure}

Let us now go beyond the highly symmetry-restricted case and consider axisymmetric, regular spacetimes. By considering $S_{\rm ax}^{(3)}$ for the rotating Hayward black hole, we can confirm its finiteness and check the Kerr and Schwarzschild results by investigating the appropriate limits of the parameters $a$ and $g$, cf.~Fig.~\ref{fig:rotHaywardint}. There, the expected finiteness of $S_{\rm ax}^{(3)}(0,\infty,0)$ is confirmed numerically for finite $g$.

These results confirm that including additional terms quadratic in the curvature tensor does not affect the finiteness of the action for regular spacetimes, as expected.

An action that is divergent on a given configuration is expected to result in destructive interference in the path integral, suppressing the contribution of that field configuration.
Thus, the results of this subsection show that a curvature-squared action could suppress axi-symmetric and spherically symmetric, singular black-hole spacetimes. At the same time, such an action remains finite for regular black-hole spacetimes, ensuring that they could contribute to the path integral. Here, we do not investigate whether such spacetimes correspond to a saddle-point of the path integral. In particular, such a question can only be meaningfully investigated within a matter-gravity path integral, going beyond the scope of this work. 

\subsection{Beyond four-derivative dynamics}
In the previous section, we found that the inclusion of the Kretschmann scalar in the action is sufficient to exclude singular black-hole spacetimes from the path-integral based on the requirement of a finite action for non-zero contributions. Following a principle of canonical power counting, this would be the first term expected to occur in the dynamics when going beyond the Einstein-Hilbert action. Yet, the microscopic gravitational action is not bound to follow canonical power counting, and the coupling of the Kretschmann scalar might vanish in a given quantum gravity setting. This motivates us to go beyond the four-derivative terms in the dynamics.

Interestingly, within the asymptotic-safety program, there are indications that the coupling associated to the invariant $K^{(5)}$, the Goroff-Sagnotti term, is irrelevant \cite{Gies:2016con}, while there might be a gravitational universality class at which other curvature-cubed terms might come with relevant couplings \cite{Kluth:2020bdv}.

We expect that an action built from any higher-order curvature invariant that does neither contain $R$ nor $R_{\mu\nu}$ will exhibit an ultraviolet divergence, when evaluated on a singular black-hole spacetime. To provide some evidence for this expectation, we consider a complete basis at order 6 in derivatives and highlight the behavior of two order-8 invariants.

For the Schwarzschild black hole, the integrals over $K^{(4,5,7,8)}$ all exhibit a divergence in the limit $r_{\rm UV} \rightarrow 0$ which follows their dimensionality:
\bea
S_{\rm sph}^{(4,5)}(r_{\rm UV}, \infty) &\sim& \frac{G_{\rm N}^3 M^3}{r_{\rm UV}^6},\label{eq:Ssph45Schw}\\
S_{\rm sph}^{(7,8)}(r_{\rm UV}, \infty) &\sim& \frac{G_{\rm N}^4 M^4}{r_{\rm UV}^9}.\label{eq:Ssph78Schw}
\eea
The horizon-detecting invariant $K^{(6)}$ features a leading divergence $S_{\rm sph}^{(6)}(r_{\rm UV}, \infty) \sim r_{\rm UV}^{-6}$ and a subleading $r_{\rm UV}^{-5}$ term.

For the Vaidya spacetime, the action function $S_{\rm sph}^{(4,5,7,8)}$ follows from the Schwarzschild results Eq.~\eqref{eq:Ssph45Schw} and Eq.~\eqref{eq:Ssph78Schw} with the substitution $M \rightarrow M(v)$. As the invariant $K^{(6)}$ is a horizon-detecting one in the Schwarzschild limit, it depends on $M'(v)$ for the Vaidya case, and the corresponding action function reads
\be
S_{\rm sph}^{(6)}(r_{\rm UV}, \infty) = \frac{24 G_{\rm N} M(v)}{r_{\rm UV}^6} \left(6 r_{\rm UV} M(v) - 10 M(v)^2 - 3 r_{\rm UV}^2 M'(v) \right).
\ee
 
To calculate the action function $S_{\rm ax}(r_{\rm UV}, r_{\rm IR}, \epsilon)$ for the Kerr spacetime, one should again be careful how to take the limits $\epsilon \rightarrow 0$ and $r_{\rm UV} \rightarrow 0$ in order to account for the presence of the ring singularity. Thus, we set $r_{\rm UV} = \gamma \epsilon$ for some constant $\gamma >0$, and then take $\epsilon \rightarrow 0$.
In this way, we obtain
\bea
S_{\rm ax}^{(4)}(\gamma \epsilon, \infty, \epsilon) &=& \frac{16}{7}G_{\rm N}^3 M^3 \Biggl(\frac{a^6-21a^4 \gamma^2+35 a^2 \gamma^4-7 \gamma^6}{(a^2+\gamma^2)^6 \epsilon^5}- \frac{1}{a^6}\Biggr) +\mathcal{O}(\epsilon),\\
S_{\rm ax}^{(5)}(\gamma \epsilon, \infty, \epsilon) &=& 2 S_{\rm ax}^{(4)},\\
S_{\rm ax}^{(6)}(\gamma \epsilon, \infty, \epsilon) &=& \frac{96}{7}G_{\rm N}^2 M^2 \Biggl( \frac{G_{\rm N }M \left(a^6+21a^4 \gamma^2-105 a^2 \gamma^4 + 35 \gamma^6 \right)}{(a^2+\gamma^2)^6 \epsilon^5} \nonumber\\
&{}&-\frac{7\gamma  \left(3a^6-7a^4 \gamma^2 - 7 a^2 \gamma^4 + 3\gamma^6\right)}{(a^2+\gamma^2)^6 \epsilon^4} -\frac{G_{\rm N} M}{a^6}\Biggr) + \mathcal{O}(\epsilon).
\eea
Finally, constructing action functions from the two dimension-eight operators we consider, we do not obtain a finite part in the action function:
\bea
S_{\rm ax}^{(7)}(\gamma \epsilon, \infty, \epsilon) &=&-\frac{128}{5}G_{\rm N}^4 M^4 \gamma \frac{\left(5a^4-10a^2 \gamma^2 +\gamma^4\right)\left(a^4-10a^2\gamma^2+5\gamma^4 \right)}{(a^2+\gamma^2 \epsilon^2)^9\epsilon^8},\\
S_{\rm ax}^{(8)}(\gamma \epsilon, \infty, \epsilon) &=& \frac{1}{20}G_{\rm N}^4 M^4 \frac{1}{a^9 \gamma^8 \epsilon^8} \Bigl(-\frac{1}{(a^2+\gamma^2)^9}
\Bigl( 1575 a^{25} \gamma + 13650 a^{23}\gamma^3 + 52290 a^{21}\gamma^5 \nonumber\\
&{}&+116010 a^{19}\gamma^7+172505 a^{17}\gamma^9 + 116100 a^{15}\gamma^{11}+300924 a^{13}\gamma^{13} + 166100 a^{11}\gamma^{15} \nonumber\\
&{}&+172505 a^9 \gamma^{17} + 116010 a^7 \gamma^{19}+52290 a^5 \gamma^{21}+13650 a^3 \gamma^{23}+1575 a \gamma^{25}\Bigr)\nonumber\\
&{}&+1575(a^8-\gamma^8\epsilon^8){\rm arccot}\left(\frac{\gamma \epsilon}{a}\right)  - 1575 (a^8-\gamma^8){\rm arctan}\left(\frac{a}{\gamma} \right) 
\Bigr).
\eea

As we expected, the inclusion of any of these curvature invariants suffices to render the action divergent and thus provides a dynamical mechanism to resolve black-hole singularities.

\section{Conclusions and outlook}\label{sec:conclusions}
In this paper, we have used a mechanism of dynamical singularity-resolution in black-hole spacetimes to motivate the inclusion of higher-order curvature invariants in the microscopic gravitational action in the Lorentzian path integral. Specifically, we have shown that the inclusion of a complete basis of invariants constructed from the Riemann tensor of mass dimension four leads to a divergent action for the Schwarzschild, Kerr and Vaidya spacetimes. These divergences are associated to large values of the curvature, i.e., to the central curvature singularity. In turn, a divergent action in the Lorentzian path integral corresponds to an infinitely fast oscillation of the quantum-mechanical phase factor and is hence expected to result in a dynamical suppression of the corresponding field configurations through destructive interference. Thus, the presence of higher-order curvature invariants constructed from the Riemann tensor, which reflect the curvature singularities of various black-hole spacetimes, provide a dynamical mechanism to resolve those singularities. Thereby, they fulfill a key requirement for a viable quantum theory of gravity. This is in contrast to the Einstein-Hilbert action, which vanishes when evaluated on these spacetimes, as these are vacuum solutions of GR. In addition to its perturbative renormalizability, its failure to provide a mechanism for dynamical singularity-resolution is a second reason to discard the Einstein-Hilbert action as a candidate for a fundamental action for gravity. Beyond GR, either Stelle's higher-derivative gravity \cite{Stelle1977} or  asymptotically safe gravity, both of which contain appropriate higher-curvature terms, could fulfill our criterion. \\

Let us make several comments regarding our results:
\begin{enumerate}
\item[i)] To dynamically suppress singular black-hole spacetimes that are vacuum solutions of Einstein equations, a local action constructed from higher powers of $R$ or $R_{\mu\nu}$ alone is insufficient and invariants built out of the Riemann tensor are required. In turn, a local action containing these further invariants may feature ghosts \footnote{These could be avoided by including appropriate functions of the curvature invariants \cite{Draper:2020bop,Platania:2020knd} or might be present, but not result in a physical inconsistency of the theory \cite{Anselmi:2018ibi,Donoghue:2019ecz}.} which would be absent if one would restrict to higher powers in $R$, only. The latter could be taken as a motivation to focus on $f(R)$-type of actions for gravity. Here, we provide evidence that $f(R)$ actions, while potentially providing physically interesting \emph{effective} descriptions of gravity, fail at a key requirement of quantum gravity, namely dynamical singularity-resolution. In some sense, therefore, the challenge of potential ghosts and the dynamical resolution of singularities appear to be two sides of the same coin.
\item[ii)] The contribution of black-hole spacetimes to the gravitational path integral has been used to argue that quantum gravity should result in the violation of global symmetries \cite{Kamionkowski:1992mf,Kallosh:1995hi,Banks:2010zn}. The argument hinges  on the presence of black-hole spacetimes in the gravitational path integral. As we have argued, this depends on the choice of dynamics. The latter is also key to understand the final state of black-hole evaporation. Our study motivates that under certain circumstances, at least the contribution of certain singular black-hole spacetimes to the gravitational path integral could vanish. This could in turn imply that quantum-gravity approaches which feature such higher-order terms in the action might not generically violate global symmetries. This might for instance include asymptotically safe gravity, where indeed results in Euclidean studies of truncated dynamics for gravity-matter systems do not provide any indications for the violation of global symmetries \cite{Eichhorn:2017eht,Eichhorn:2020sbo}.
\end{enumerate}

Our work motivates several potential extensions:\\
 First, our study can be extended to include additional singular black-hole spacetimes, in particular additional models of spherical and non-spherical collapse, in order to investigate whether our proposed mechanism for dynamical singularity-resolution also removes those spacetimes from the path integral. Since perturbations of black-hole spacetimes, i.e., quasinormal modes, do not affect the presence of the singularity, our present study already extends directly to perturbed Kerr and Schwarzschild spacetimes.\\
 Second, there is an interesting question to explore in a semi-classical framework: By requiring that regular black-hole spacetimes correspond to a saddle-point of the action, one might find a principle that further distinguishes between the various possible curvature terms. \\
Third, given a candidate for the microscopic action $S$ in the Lorentzian path integral, motivated by singularity-resolution, one may next ask whether this action provides a different conclusion to the viability of the Hartle-Hawking wavefunction, using the methods in \cite{Feldbrugge:2017kzv,Feldbrugge:2017fcc}.\\
Fourth, changing the underlying symmetries in the path integral, e.g., to foliation-preserving diffeomorphisms, as, e.g., in Horava-Lifshitz gravity \cite{Horava2009} (see \cite{Steinwachs:2020jkj} for a recent review), could require different invariants to be included for singularity-resolution. In this spirit, a comparison of diverse quantum-field-theoretic proposals for gravity which differ in a) the underlying (gauge) symmetries and/or b) the choice of fields, see, e.g., \cite{Daum:2013fu,Harst:2012ni,Knorr:2020bjm,deBrito:2020rwu} for proposals which are classically equivalent to GR, could be illuminating. In particular, demanding destructive interference between singular field configurations could provide a principle to distinguish between different candidate theories.

\vspace{6pt}

\funding{A.~E.~is supported by a grant by Villum Fonden under grant (29405) and a grant by the DFG under grant no.~Ei/1037-1. J.~N.~B.~is supported by the German Academic Scholarship Foundation.
}

\acknowledgments{We thank Jean-Luc Lehners and Kelly Stelle for discussions on \cite{Lehners:2019ibe} and Aaron Held for discussions on curvature invariants.}

\reftitle{References}

\externalbibliography{yes}
\bibliography{References}

\begin{thebibliography}{-------}
\providecommand{\natexlab}[1]{#1}

\bibitem[Abbott \em{et~al.}(2016)Abbott et~al.]{Abbott:2016blz}
Abbott, B.P.; others.
\newblock {Observation of Gravitational Waves from a Binary Black Hole Merger}.
\newblock {\em Phys. Rev. Lett.} {\bf 2016}, {\em 116},~061102,
  \href{http://xxx.lanl.gov/abs/1602.03837}{{\normalfont
  [arXiv:gr-qc/1602.03837]}}.
\newblock
  doi:{\changeurlcolor{black}\href{https://doi.org/10.1103/PhysRevLett.116.061102}{\detokenize{10.1103/PhysRevLett.116.061102}}}.

\bibitem[Ghez \em{et~al.}(2008)Ghez et~al.]{Ghez:2008ms}
Ghez, A.M.; others.
\newblock {Measuring Distance and Properties of the Milky Way's Central
  Supermassive Black Hole with Stellar Orbits}.
\newblock {\em Astrophys. J.} {\bf 2008}, {\em 689},~1044--1062,
  \href{http://xxx.lanl.gov/abs/0808.2870}{{\normalfont
  [arXiv:astro-ph/0808.2870]}}.
\newblock
  doi:{\changeurlcolor{black}\href{https://doi.org/10.1086/592738}{\detokenize{10.1086/592738}}}.

\bibitem[Akiyama \em{et~al.}(2019)Akiyama et~al.]{Akiyama:2019cqa}
Akiyama, K.; others.
\newblock {First M87 Event Horizon Telescope Results. I. The Shadow of the
  Supermassive Black Hole}.
\newblock {\em Astrophys. J.} {\bf 2019}, {\em 875},~L1,
  \href{http://xxx.lanl.gov/abs/1906.11238}{{\normalfont
  [arXiv:astro-ph.GA/1906.11238]}}.
\newblock
  doi:{\changeurlcolor{black}\href{https://doi.org/10.3847/2041-8213/ab0ec7}{\detokenize{10.3847/2041-8213/ab0ec7}}}.

\bibitem[Lehners and Stelle(2019)]{Lehners:2019ibe}
Lehners, J.L.; Stelle, K.
\newblock {A Safe Beginning for the Universe?}
\newblock {\em Phys. Rev. D} {\bf 2019}, {\em 100},~083540,
  \href{http://xxx.lanl.gov/abs/1909.01169}{{\normalfont
  [arXiv:hep-th/1909.01169]}}.
\newblock
  doi:{\changeurlcolor{black}\href{https://doi.org/10.1103/PhysRevD.100.083540}{\detokenize{10.1103/PhysRevD.100.083540}}}.

\bibitem[Calmet \em{et~al.}(2008)Calmet, Hsu, and Reeb]{Calmet:2008tn}
Calmet, X.; Hsu, S.D.; Reeb, D.
\newblock {Quantum gravity at a TeV and the renormalization of Newton's
  constant}.
\newblock {\em Phys. Rev. D} {\bf 2008}, {\em 77},~125015,
  \href{http://xxx.lanl.gov/abs/0803.1836}{{\normalfont
  [arXiv:hep-th/0803.1836]}}.
\newblock
  doi:{\changeurlcolor{black}\href{https://doi.org/10.1103/PhysRevD.77.125015}{\detokenize{10.1103/PhysRevD.77.125015}}}.

\bibitem[Hoyle \em{et~al.}(2004)Hoyle, Kapner, Heckel, Adelberger, Gundlach,
  Schmidt, and Swanson]{Hoyle:2004cw}
Hoyle, C.; Kapner, D.; Heckel, B.R.; Adelberger, E.; Gundlach, J.; Schmidt, U.;
  Swanson, H.
\newblock {Sub-millimeter tests of the gravitational inverse-square law}.
\newblock {\em Phys. Rev. D} {\bf 2004}, {\em 70},~042004,
  \href{http://xxx.lanl.gov/abs/hep-ph/0405262}{{\normalfont
  [hep-ph/0405262]}}.
\newblock
  doi:{\changeurlcolor{black}\href{https://doi.org/10.1103/PhysRevD.70.042004}{\detokenize{10.1103/PhysRevD.70.042004}}}.

\bibitem['t~Hooft and Veltman(1974)]{tHooftVeltman1974}
't~Hooft, G.; Veltman, M.J.G.
\newblock {One loop divergencies in the theory of gravitation}.
\newblock {\em Ann. Inst. H. Poincare Phys. Theor.} {\bf 1974}, {\em
  A20},~69--94.

\bibitem[Goroff and Sagnotti(1986)]{Goroff1986}
Goroff, M.H.; Sagnotti, A.
\newblock {The Ultraviolet Behavior of Einstein Gravity}.
\newblock {\em Nucl. Phys. B} {\bf 1986}, {\em 266},~709--736.
\newblock
  doi:{\changeurlcolor{black}\href{https://doi.org/10.1016/0550-3213(86)90193-8}{\detokenize{10.1016/0550-3213(86)90193-8}}}.

\bibitem[van~de Ven(1992)]{VandeVen1991}
van~de Ven, A.E.M.
\newblock {Two loop quantum gravity}.
\newblock {\em Nucl. Phys. B} {\bf 1992}, {\em 378},~309--366.
\newblock
  doi:{\changeurlcolor{black}\href{https://doi.org/10.1016/0550-3213(92)90011-Y}{\detokenize{10.1016/0550-3213(92)90011-Y}}}.

\bibitem[Donoghue(1994)]{Donoghue1994}
Donoghue, J.F.
\newblock {Leading quantum correction to the Newtonian potential}.
\newblock {\em Phys. Rev. Lett.} {\bf 1994}, {\em 72},~2996--2999.
\newblock
  doi:{\changeurlcolor{black}\href{https://doi.org/10.1103/PhysRevLett.72.2996}{\detokenize{10.1103/PhysRevLett.72.2996}}}.

\bibitem[Donoghue(1995)]{Donoghue:1995cz}
Donoghue, J.F.
\newblock {Introduction to the effective field theory description of gravity}.
\newblock  {Advanced School on Effective Theories},  1995,
  \href{http://xxx.lanl.gov/abs/gr-qc/9512024}{{\normalfont [gr-qc/9512024]}}.

\bibitem[Hartle and Hawking(1987)]{Hartle:1983ai}
Hartle, J.; Hawking, S.
\newblock {Wave Function of the Universe}.
\newblock {\em Adv. Ser. Astrophys. Cosmol.} {\bf 1987}, {\em 3},~174--189.
\newblock
  doi:{\changeurlcolor{black}\href{https://doi.org/10.1103/PhysRevD.28.2960}{\detokenize{10.1103/PhysRevD.28.2960}}}.

\bibitem[Feldbrugge \em{et~al.}(2017)Feldbrugge, Lehners, and
  Turok]{Feldbrugge:2017fcc}
Feldbrugge, J.; Lehners, J.L.; Turok, N.
\newblock {No smooth beginning for spacetime}.
\newblock {\em Phys. Rev. Lett.} {\bf 2017}, {\em 119},~171301,
  \href{http://xxx.lanl.gov/abs/1705.00192}{{\normalfont
  [arXiv:hep-th/1705.00192]}}.
\newblock
  doi:{\changeurlcolor{black}\href{https://doi.org/10.1103/PhysRevLett.119.171301}{\detokenize{10.1103/PhysRevLett.119.171301}}}.

\bibitem[Halliwell \em{et~al.}(2019)Halliwell, Hartle, and
  Hertog]{Halliwell:2018ejl}
Halliwell, J.J.; Hartle, J.B.; Hertog, T.
\newblock {What is the No-Boundary Wave Function of the Universe?}
\newblock {\em Phys. Rev. D} {\bf 2019}, {\em 99},~043526,
  \href{http://xxx.lanl.gov/abs/1812.01760}{{\normalfont
  [arXiv:hep-th/1812.01760]}}.
\newblock
  doi:{\changeurlcolor{black}\href{https://doi.org/10.1103/PhysRevD.99.043526}{\detokenize{10.1103/PhysRevD.99.043526}}}.

\bibitem[Stelle(1977)]{Stelle1977}
Stelle, K.S.
\newblock {Renormalization of Higher Derivative Quantum Gravity}.
\newblock {\em Phys. Rev.} {\bf 1977}, {\em D16},~953--969.
\newblock
  doi:{\changeurlcolor{black}\href{https://doi.org/10.1103/PhysRevD.16.953}{\detokenize{10.1103/PhysRevD.16.953}}}.

\bibitem[Fradkin and Tseytlin(1982)]{Fradkin1982}
Fradkin, E.S.; Tseytlin, A.A.
\newblock {Renormalizable asymptotically free quantum theory of gravity}.
\newblock {\em Nucl. Phys. B} {\bf 1982}, {\em 201},~469--491.
\newblock
  doi:{\changeurlcolor{black}\href{https://doi.org/10.1016/0550-3213(82)90444-8}{\detokenize{10.1016/0550-3213(82)90444-8}}}.

\bibitem[Salvio(2018)]{Salvio:2018crh}
Salvio, A.
\newblock {Quadratic Gravity}.
\newblock {\em Front. in Phys.} {\bf 2018}, {\em 6},~77,
  \href{http://xxx.lanl.gov/abs/1804.09944}{{\normalfont
  [arXiv:hep-th/1804.09944]}}.
\newblock
  doi:{\changeurlcolor{black}\href{https://doi.org/10.3389/fphy.2018.00077}{\detokenize{10.3389/fphy.2018.00077}}}.

\bibitem[Anselmi and Piva(2018)]{Anselmi:2018ibi}
Anselmi, D.; Piva, M.
\newblock {The Ultraviolet Behavior of Quantum Gravity}.
\newblock {\em JHEP} {\bf 2018}, {\em 05},~027,
  \href{http://xxx.lanl.gov/abs/1803.07777}{{\normalfont
  [arXiv:hep-th/1803.07777]}}.
\newblock
  doi:{\changeurlcolor{black}\href{https://doi.org/10.1007/JHEP05(2018)027}{\detokenize{10.1007/JHEP05(2018)027}}}.

\bibitem[Donoghue and Menezes(2019)]{Donoghue:2019ecz}
Donoghue, J.F.; Menezes, G.
\newblock {Arrow of Causality and Quantum Gravity}.
\newblock {\em Phys. Rev. Lett.} {\bf 2019}, {\em 123},~171601,
  \href{http://xxx.lanl.gov/abs/1908.04170}{{\normalfont
  [arXiv:hep-th/1908.04170]}}.
\newblock
  doi:{\changeurlcolor{black}\href{https://doi.org/10.1103/PhysRevLett.123.171601}{\detokenize{10.1103/PhysRevLett.123.171601}}}.

\bibitem[Weinberg(1979)]{Weinberg1979}
Weinberg, S.
\newblock {Ultraviolet divergencies in quantum theories of gravitation}. In
  {\em General Relativity: An Einstein Centenary Survey};  1979; pp. 790--831.

\bibitem[Reuter(1998)]{Reuter:1996cp}
Reuter, M.
\newblock {Nonperturbative Evolution Equation for Quantum Gravity}.
\newblock {\em Phys. Rev.} {\bf 1998}, {\em D57},~971--985,
  \href{http://xxx.lanl.gov/abs/hep-th/9605030}{{\normalfont
  [hep-th/9605030]}}.
\newblock
  doi:{\changeurlcolor{black}\href{https://doi.org/10.1103/PhysRevD.57.971}{\detokenize{10.1103/PhysRevD.57.971}}}.

\bibitem[Lauscher and Reuter(2002)]{Lauscher:2002sq}
Lauscher, O.; Reuter, M.
\newblock {Flow equation of quantum Einstein gravity in a higher derivative
  truncation}.
\newblock {\em Phys. Rev. D} {\bf 2002}, {\em 66},~025026,
  \href{http://xxx.lanl.gov/abs/hep-th/0205062}{{\normalfont
  [hep-th/0205062]}}.
\newblock
  doi:{\changeurlcolor{black}\href{https://doi.org/10.1103/PhysRevD.66.025026}{\detokenize{10.1103/PhysRevD.66.025026}}}.

\bibitem[Machado and Saueressig(2008)]{Machado:2007ea}
Machado, P.F.; Saueressig, F.
\newblock {On the renormalization group flow of f(R)-gravity}.
\newblock {\em Phys. Rev. D} {\bf 2008}, {\em 77},~124045,
  \href{http://xxx.lanl.gov/abs/0712.0445}{{\normalfont
  [arXiv:hep-th/0712.0445]}}.
\newblock
  doi:{\changeurlcolor{black}\href{https://doi.org/10.1103/PhysRevD.77.124045}{\detokenize{10.1103/PhysRevD.77.124045}}}.

\bibitem[Codello \em{et~al.}(2009)Codello, Percacci, and
  Rahmede]{Codello:2008vh}
Codello, A.; Percacci, R.; Rahmede, C.
\newblock {Investigating the Ultraviolet Properties of Gravity with a Wilsonian
  Renormalization Group Equation}.
\newblock {\em Annals Phys.} {\bf 2009}, {\em 324},~414--469,
  \href{http://xxx.lanl.gov/abs/0805.2909}{{\normalfont
  [arXiv:hep-th/0805.2909]}}.
\newblock
  doi:{\changeurlcolor{black}\href{https://doi.org/10.1016/j.aop.2008.08.008}{\detokenize{10.1016/j.aop.2008.08.008}}}.

\bibitem[Benedetti \em{et~al.}(2009)Benedetti, Machado, and
  Saueressig]{Benedetti:2009rx}
Benedetti, D.; Machado, P.F.; Saueressig, F.
\newblock {Asymptotic safety in higher-derivative gravity}.
\newblock {\em Mod. Phys. Lett.} {\bf 2009}, {\em A24},~2233--2241,
  \href{http://xxx.lanl.gov/abs/0901.2984}{{\normalfont
  [arXiv:hep-th/0901.2984]}}.
\newblock
  doi:{\changeurlcolor{black}\href{https://doi.org/10.1142/S0217732309031521}{\detokenize{10.1142/S0217732309031521}}}.

\bibitem[Dietz and Morris(2013)]{Dietz:2012ic}
Dietz, J.A.; Morris, T.R.
\newblock {Asymptotic safety in the f(R) approximation}.
\newblock {\em JHEP} {\bf 2013}, {\em 01},~108,
  \href{http://xxx.lanl.gov/abs/1211.0955}{{\normalfont
  [arXiv:hep-th/1211.0955]}}.
\newblock
  doi:{\changeurlcolor{black}\href{https://doi.org/10.1007/JHEP01(2013)108}{\detokenize{10.1007/JHEP01(2013)108}}}.

\bibitem[Benedetti and Caravelli(2012)]{Benedetti:2012dx}
Benedetti, D.; Caravelli, F.
\newblock {The Local potential approximation in quantum gravity}.
\newblock {\em JHEP} {\bf 2012}, {\em 06},~017,
  \href{http://xxx.lanl.gov/abs/1204.3541}{{\normalfont
  [arXiv:hep-th/1204.3541]}}.
\newblock [Erratum: JHEP 10, 157 (2012)],
  doi:{\changeurlcolor{black}\href{https://doi.org/10.1007/JHEP06(2012)017}{\detokenize{10.1007/JHEP06(2012)017}}}.

\bibitem[Falls \em{et~al.}(2013)Falls, Litim, Nikolakopoulos, and
  Rahmede]{Falls:2013bv}
Falls, K.; Litim, D.F.; Nikolakopoulos, K.; Rahmede, C.
\newblock {A bootstrap towards asymptotic safety}.
\newblock {\em arXiv preprints} {\bf 2013},
  \href{http://xxx.lanl.gov/abs/1301.4191}{{\normalfont
  [arXiv:hep-th/1301.4191]}}.

\bibitem[Ohta \em{et~al.}(2016)Ohta, Percacci, and Vacca]{Ohta:2015fcu}
Ohta, N.; Percacci, R.; Vacca, G.P.
\newblock {Renormalization Group Equation and scaling solutions for f(R)
  gravity in exponential parametrization}.
\newblock {\em Eur. Phys. J. C} {\bf 2016}, {\em 76},~46,
  \href{http://xxx.lanl.gov/abs/1511.09393}{{\normalfont
  [arXiv:hep-th/1511.09393]}}.
\newblock
  doi:{\changeurlcolor{black}\href{https://doi.org/10.1140/epjc/s10052-016-3895-1}{\detokenize{10.1140/epjc/s10052-016-3895-1}}}.

\bibitem[Demmel \em{et~al.}(2015)Demmel, Saueressig, and
  Zanusso]{Demmel:2015oqa}
Demmel, M.; Saueressig, F.; Zanusso, O.
\newblock {A proper fixed functional for four-dimensional Quantum Einstein
  Gravity}.
\newblock {\em JHEP} {\bf 2015}, {\em 08},~113,
  \href{http://xxx.lanl.gov/abs/1504.07656}{{\normalfont
  [arXiv:hep-th/1504.07656]}}.
\newblock
  doi:{\changeurlcolor{black}\href{https://doi.org/10.1007/JHEP08(2015)113}{\detokenize{10.1007/JHEP08(2015)113}}}.

\bibitem[Christiansen \em{et~al.}(2018)Christiansen, Falls, Pawlowski, and
  Reichert]{Christiansen:2017bsy}
Christiansen, N.; Falls, K.; Pawlowski, J.M.; Reichert, M.
\newblock {Curvature dependence of quantum gravity}.
\newblock {\em Phys. Rev.} {\bf 2018}, {\em D97},~046007,
  \href{http://xxx.lanl.gov/abs/1711.09259}{{\normalfont
  [arXiv:hep-th/1711.09259]}}.
\newblock
  doi:{\changeurlcolor{black}\href{https://doi.org/10.1103/PhysRevD.97.046007}{\detokenize{10.1103/PhysRevD.97.046007}}}.

\bibitem[Gonzalez-Martin \em{et~al.}(2017)Gonzalez-Martin, Morris, and
  Slade]{Gonzalez-Martin:2017gza}
Gonzalez-Martin, S.; Morris, T.R.; Slade, Z.H.
\newblock {Asymptotic solutions in asymptotic safety}.
\newblock {\em Phys. Rev. D} {\bf 2017}, {\em 95},~106010,
  \href{http://xxx.lanl.gov/abs/1704.08873}{{\normalfont
  [arXiv:hep-th/1704.08873]}}.
\newblock
  doi:{\changeurlcolor{black}\href{https://doi.org/10.1103/PhysRevD.95.106010}{\detokenize{10.1103/PhysRevD.95.106010}}}.

\bibitem[Falls \em{et~al.}(2018)Falls, King, Litim, Nikolakopoulos, and
  Rahmede]{Falls:2017lst}
Falls, K.; King, C.R.; Litim, D.F.; Nikolakopoulos, K.; Rahmede, C.
\newblock {Asymptotic safety of quantum gravity beyond Ricci scalars}.
\newblock {\em Phys. Rev. D} {\bf 2018}, {\em 97},~086006,
  \href{http://xxx.lanl.gov/abs/1801.00162}{{\normalfont
  [arXiv:hep-th/1801.00162]}}.
\newblock
  doi:{\changeurlcolor{black}\href{https://doi.org/10.1103/PhysRevD.97.086006}{\detokenize{10.1103/PhysRevD.97.086006}}}.

\bibitem[De~Brito \em{et~al.}(2018)De~Brito, Ohta, Pereira, Tomaz, and
  Yamada]{deBrito:2018jxt}
De~Brito, G.P.; Ohta, N.; Pereira, A.D.; Tomaz, A.A.; Yamada, M.
\newblock {Asymptotic safety and field parametrization dependence in the $f(R)$
  truncation}.
\newblock {\em Phys. Rev. D} {\bf 2018}, {\em 98},~026027,
  \href{http://xxx.lanl.gov/abs/1805.09656}{{\normalfont
  [arXiv:hep-th/1805.09656]}}.
\newblock
  doi:{\changeurlcolor{black}\href{https://doi.org/10.1103/PhysRevD.98.026027}{\detokenize{10.1103/PhysRevD.98.026027}}}.

\bibitem[Falls \em{et~al.}(2020)Falls, Ohta, and Percacci]{Falls:2020qhj}
Falls, K.; Ohta, N.; Percacci, R.
\newblock {Towards the determination of the dimension of the critical surface
  in asymptotically safe gravity}.
\newblock {\em Phys. Lett. B} {\bf 2020}, {\em 810},~135773,
  \href{http://xxx.lanl.gov/abs/2004.04126}{{\normalfont
  [arXiv:hep-th/2004.04126]}}.
\newblock
  doi:{\changeurlcolor{black}\href{https://doi.org/10.1016/j.physletb.2020.135773}{\detokenize{10.1016/j.physletb.2020.135773}}}.

\bibitem[Eichhorn(2015)]{Eichhorn:2015bna}
Eichhorn, A.
\newblock {The Renormalization Group flow of unimodular f(R) gravity}.
\newblock {\em JHEP} {\bf 2015}, {\em 04},~096,
  \href{http://xxx.lanl.gov/abs/1501.05848}{{\normalfont
  [arXiv:gr-qc/1501.05848]}}.
\newblock
  doi:{\changeurlcolor{black}\href{https://doi.org/10.1007/JHEP04(2015)096}{\detokenize{10.1007/JHEP04(2015)096}}}.

\bibitem[Percacci(2017)]{Percacci2017}
Percacci, R.
\newblock {\em {An Introduction to Covariant Quantum Gravity and Asymptotic
  Safety}}; Vol.~3, {\em 100 Years of General Relativity}, World Scientific:
  Singapore,  2017.
\newblock
  doi:{\changeurlcolor{black}\href{https://doi.org/10.1142/10369}{\detokenize{10.1142/10369}}}.

\bibitem[Eichhorn(2019)]{Eichhorn:2018yfc}
Eichhorn, A.
\newblock {An asymptotically safe guide to quantum gravity and matter}.
\newblock {\em Frontiers in Astronomy and Space Sciences} {\bf 2019}, {\em
  5},~47,  \href{http://xxx.lanl.gov/abs/1810.07615}{{\normalfont
  [arXiv:hep-th/1810.07615]}}.
\newblock
  doi:{\changeurlcolor{black}\href{https://doi.org/10.3389/fspas.2018.00047}{\detokenize{10.3389/fspas.2018.00047}}}.

\bibitem[Reuter and Saueressig(2019)]{ReuterSaueressig2019}
Reuter, M.; Saueressig, F.
\newblock {\em {Quantum Gravity and the Functional Renormalization Group}: {The
  Road towards Asymptotic Safety}}; Cambridge University Press: Cambridge,
  2019.

\bibitem[Manrique \em{et~al.}(2011)Manrique, Rechenberger, and
  Saueressig]{Manrique:2011jc}
Manrique, E.; Rechenberger, S.; Saueressig, F.
\newblock {Asymptotically Safe Lorentzian Gravity}.
\newblock {\em Phys.Rev.Lett.} {\bf 2011}, {\em 106},~251302,
  \href{http://xxx.lanl.gov/abs/1102.5012}{{\normalfont
  [arXiv:hep-th/1102.5012]}}.
\newblock
  doi:{\changeurlcolor{black}\href{https://doi.org/10.1103/PhysRevLett.106.251302}{\detokenize{10.1103/PhysRevLett.106.251302}}}.

\bibitem[Bonanno \em{et~al.}(2020)Bonanno, Eichhorn, Gies, Pawlowski, Percacci,
  Reuter, Saueressig, and Vacca]{Bonanno:2020bil}
Bonanno, A.; Eichhorn, A.; Gies, H.; Pawlowski, J.M.; Percacci, R.; Reuter, M.;
  Saueressig, F.; Vacca, G.P.
\newblock {Critical reflections on asymptotically safe gravity} {\bf 2020}.
\newblock  \href{http://xxx.lanl.gov/abs/2004.06810}{{\normalfont
  [arXiv:gr-qc/2004.06810]}}.

\bibitem[Dupuis \em{et~al.}(2020)Dupuis, Canet, Eichhorn, Metzner, Pawlowski,
  Tissier, and Wschebor]{Dupuis:2020fhh}
Dupuis, N.; Canet, L.; Eichhorn, A.; Metzner, W.; Pawlowski, J.; Tissier, M.;
  Wschebor, N.
\newblock {The nonperturbative functional renormalization group and its
  applications} {\bf 2020}.
\newblock  \href{http://xxx.lanl.gov/abs/2006.04853}{{\normalfont
  [arXiv:cond-mat.stat-mech/2006.04853]}}.

\bibitem[Manrique and Reuter(2009)]{Manrique:2008zw}
Manrique, E.; Reuter, M.
\newblock {Bare Action and Regularized Functional Integral of Asymptotically
  Safe Quantum Gravity}.
\newblock {\em Phys. Rev. D} {\bf 2009}, {\em 79},~025008,
  \href{http://xxx.lanl.gov/abs/0811.3888}{{\normalfont
  [arXiv:hep-th/0811.3888]}}.
\newblock
  doi:{\changeurlcolor{black}\href{https://doi.org/10.1103/PhysRevD.79.025008}{\detokenize{10.1103/PhysRevD.79.025008}}}.

\bibitem[Manrique and Reuter(2011)]{Manrique:2009tj}
Manrique, E.; Reuter, M.
\newblock {Bare versus Effective Fixed Point Action in Asymptotic Safety: The
  Reconstruction Problem}.
\newblock {\em PoS} {\bf 2011}, {\em CLAQG08},~001,
  \href{http://xxx.lanl.gov/abs/0905.4220}{{\normalfont
  [arXiv:hep-th/0905.4220]}}.
\newblock
  doi:{\changeurlcolor{black}\href{https://doi.org/10.22323/1.079.0001}{\detokenize{10.22323/1.079.0001}}}.

\bibitem[Morris and Slade(2015)]{Morris:2015oca}
Morris, T.R.; Slade, Z.H.
\newblock {Solutions to the reconstruction problem in asymptotic safety}.
\newblock {\em JHEP} {\bf 2015}, {\em 11},~094,
  \href{http://xxx.lanl.gov/abs/1507.08657}{{\normalfont
  [arXiv:hep-th/1507.08657]}}.
\newblock
  doi:{\changeurlcolor{black}\href{https://doi.org/10.1007/JHEP11(2015)094}{\detokenize{10.1007/JHEP11(2015)094}}}.

\bibitem[Narlikar and Karmarkar(1949)]{Narlikar}
Narlikar, V.V.; Karmarkar, K.R.
\newblock The scalar invariants of a general gravitational metric.
\newblock {\em Proceedings of the Indian Academy of Sciences - Section A} {\bf
  1949}, {\em 29},~92.
\newblock
  doi:{\changeurlcolor{black}\href{https://doi.org/10.1007/BF03171357}{\detokenize{10.1007/BF03171357}}}.

\bibitem[Overduin \em{et~al.}(2020)Overduin, Coplan, Wilcomb, and
  Henry]{Overduin:2020aiq}
Overduin, J.; Coplan, M.; Wilcomb, K.; Henry, R.C.
\newblock {Curvature Invariants for Charged and RotatingBlack Holes}.
\newblock {\em Universe} {\bf 2020}, {\em 6},~22.
\newblock
  doi:{\changeurlcolor{black}\href{https://doi.org/10.3390/universe6020022}{\detokenize{10.3390/universe6020022}}}.

\bibitem[Fulling \em{et~al.}(1992)Fulling, King, Wybourne, and
  Cummins]{Fulling1992}
Fulling, S.A.; King, R.C.; Wybourne, B.G.; Cummins, C.J.
\newblock {Normal forms for tensor polynomials 1: The Riemann tensor}.
\newblock {\em Class. Quant. Grav.} {\bf 1992}, {\em 9},~1151--1197.
\newblock
  doi:{\changeurlcolor{black}\href{https://doi.org/10.1088/0264-9381/9/5/003}{\detokenize{10.1088/0264-9381/9/5/003}}}.

\bibitem[Lu \em{et~al.}(2015)Lu, Perkins, Pope, and Stelle]{Lu:2015cqa}
Lu, H.; Perkins, A.; Pope, C.N.; Stelle, K.S.
\newblock {Black Holes in Higher-Derivative Gravity}.
\newblock {\em Phys. Rev. Lett.} {\bf 2015}, {\em 114},~171601,
  \href{http://xxx.lanl.gov/abs/1502.01028}{{\normalfont
  [arXiv:hep-th/1502.01028]}}.
\newblock
  doi:{\changeurlcolor{black}\href{https://doi.org/10.1103/PhysRevLett.114.171601}{\detokenize{10.1103/PhysRevLett.114.171601}}}.

\bibitem[Lü \em{et~al.}(2015)Lü, Perkins, Pope, and Stelle]{Lu:2015psa}
Lü, H.; Perkins, A.; Pope, C.N.; Stelle, K.S.
\newblock {Spherically Symmetric Solutions in Higher-Derivative Gravity}.
\newblock {\em Phys. Rev.} {\bf 2015}, {\em D92},~124019,
  \href{http://xxx.lanl.gov/abs/1508.00010}{{\normalfont
  [arXiv:hep-th/1508.00010]}}.
\newblock
  doi:{\changeurlcolor{black}\href{https://doi.org/10.1103/PhysRevD.92.124019}{\detokenize{10.1103/PhysRevD.92.124019}}}.

\bibitem[Abdelqader and Lake(2015)]{Abdelqader:2014vaa}
Abdelqader, M.; Lake, K.
\newblock {Invariant characterization of the Kerr spacetime: Locating the
  horizon and measuring the mass and spin of rotating black holes using
  curvature invariants}.
\newblock {\em Phys. Rev. D} {\bf 2015}, {\em 91},~084017,
  \href{http://xxx.lanl.gov/abs/1412.8757}{{\normalfont
  [arXiv:gr-qc/1412.8757]}}.
\newblock
  doi:{\changeurlcolor{black}\href{https://doi.org/10.1103/PhysRevD.91.084017}{\detokenize{10.1103/PhysRevD.91.084017}}}.

\bibitem[Visser(2007)]{Visser:2007fj}
Visser, M.
\newblock {The Kerr spacetime: A Brief introduction}.
\newblock  {Kerr Fest: Black Holes in Astrophysics, General Relativity and
  Quantum Gravity},  2007,
  \href{http://xxx.lanl.gov/abs/0706.0622}{{\normalfont
  [arXiv:gr-qc/0706.0622]}}.

\bibitem[Vaidya(1951)]{Vaidya1951}
Vaidya, P.C.
\newblock {Nonstatic Solutions of Einstein's Field Equations for Spheres of
  Fluids Radiating Energy}.
\newblock {\em Phys. Rev.} {\bf 1951}, {\em 83},~10--17.
\newblock
  doi:{\changeurlcolor{black}\href{https://doi.org/10.1103/PhysRev.83.10}{\detokenize{10.1103/PhysRev.83.10}}}.

\bibitem[Vaidya(1953)]{Vaidya1953}
Vaidya, P.C.
\newblock {Newtonian Time in General Relativity}.
\newblock {\em Nature} {\bf 1953}, {\em 171},~260--261.
\newblock
  doi:{\changeurlcolor{black}\href{https://doi.org/10.1038/171260a0}{\detokenize{10.1038/171260a0}}}.

\bibitem[Vaidya(1966)]{Vaidya1966}
Vaidya, P.C.
\newblock {An Analytical Solution for Gravitational Collapse with Radiation}.
\newblock {\em Astrophys.\ J.} {\bf 1966}, {\em 144},~943.
\newblock
  doi:{\changeurlcolor{black}\href{https://doi.org/10.1086/148692}{\detokenize{10.1086/148692}}}.

\bibitem[Kuroda(1984)]{Kuroda1984}
Kuroda, Y.
\newblock {Naked Singularities in the Vaidya Spacetime}.
\newblock {\em Progress of Theoretical Physics} {\bf 1984}, {\em 72},~63--72.
\newblock
  doi:{\changeurlcolor{black}\href{https://doi.org/10.1143/PTP.72.63}{\detokenize{10.1143/PTP.72.63}}}.

\bibitem[Bonanno and Reuter(2000)]{Bonanno2000}
Bonanno, A.; Reuter, M.
\newblock {Renormalization group improved black hole space-times}.
\newblock {\em Phys. Rev.} {\bf 2000}, {\em D62},~043008.
\newblock
  doi:{\changeurlcolor{black}\href{https://doi.org/10.1103/PhysRevD.62.043008}{\detokenize{10.1103/PhysRevD.62.043008}}}.

\bibitem[Ashtekar and Bojowald(2006)]{Ashtekar:2005qt}
Ashtekar, A.; Bojowald, M.
\newblock {Quantum geometry and the Schwarzschild singularity}.
\newblock {\em Class. Quant. Grav.} {\bf 2006}, {\em 23},~391--411,
  \href{http://xxx.lanl.gov/abs/gr-qc/0509075}{{\normalfont [gr-qc/0509075]}}.
\newblock
  doi:{\changeurlcolor{black}\href{https://doi.org/10.1088/0264-9381/23/2/008}{\detokenize{10.1088/0264-9381/23/2/008}}}.

\bibitem[Modesto(2006)]{Modesto:2005zm}
Modesto, L.
\newblock {Loop quantum black hole}.
\newblock {\em Class. Quant. Grav.} {\bf 2006}, {\em 23},~5587--5602,
  \href{http://xxx.lanl.gov/abs/gr-qc/0509078}{{\normalfont [gr-qc/0509078]}}.
\newblock
  doi:{\changeurlcolor{black}\href{https://doi.org/10.1088/0264-9381/23/18/006}{\detokenize{10.1088/0264-9381/23/18/006}}}.

\bibitem[Bonanno and Reuter(2006)]{Bonanno:2006eu}
Bonanno, A.; Reuter, M.
\newblock {Spacetime structure of an evaporating black hole in quantum
  gravity}.
\newblock {\em Phys. Rev. D} {\bf 2006}, {\em 73},~083005,
  \href{http://xxx.lanl.gov/abs/hep-th/0602159}{{\normalfont
  [hep-th/0602159]}}.
\newblock
  doi:{\changeurlcolor{black}\href{https://doi.org/10.1103/PhysRevD.73.083005}{\detokenize{10.1103/PhysRevD.73.083005}}}.

\bibitem[Falls \em{et~al.}(2012)Falls, Litim, and Raghuraman]{Falls:2010he}
Falls, K.; Litim, D.F.; Raghuraman, A.
\newblock {Black Holes and Asymptotically Safe Gravity}.
\newblock {\em Int. J. Mod. Phys.} {\bf 2012}, {\em A27},~1250019,
  \href{http://xxx.lanl.gov/abs/1002.0260}{{\normalfont
  [arXiv:hep-th/1002.0260]}}.
\newblock
  doi:{\changeurlcolor{black}\href{https://doi.org/10.1142/S0217751X12500194}{\detokenize{10.1142/S0217751X12500194}}}.

\bibitem[Held \em{et~al.}(2019)Held, Gold, and Eichhorn]{Held2019}
Held, A.; Gold, R.; Eichhorn, A.
\newblock {Asymptotic safety casts its shadow}.
\newblock {\em JCAP} {\bf 2019}, {\em 1906},~029.
\newblock
  doi:{\changeurlcolor{black}\href{https://doi.org/10.1088/1475-7516/2019/06/029}{\detokenize{10.1088/1475-7516/2019/06/029}}}.

\bibitem[Platania(2019)]{Platania:2019kyx}
Platania, A.
\newblock {Dynamical renormalization of black-hole spacetimes}.
\newblock {\em Eur. Phys. J. C} {\bf 2019}, {\em 79},~470,
  \href{http://xxx.lanl.gov/abs/1903.10411}{{\normalfont
  [arXiv:gr-qc/1903.10411]}}.
\newblock
  doi:{\changeurlcolor{black}\href{https://doi.org/10.1140/epjc/s10052-019-6990-2}{\detokenize{10.1140/epjc/s10052-019-6990-2}}}.

\bibitem[Faraoni and Giusti(2020)]{Faraoni:2020stz}
Faraoni, V.; Giusti, A.
\newblock {Unsettling physics in the quantum-corrected Schwarzschild black
  hole}.
\newblock {\em Symmetry} {\bf 2020}, {\em 12},~1264,
  \href{http://xxx.lanl.gov/abs/2006.12577}{{\normalfont
  [arXiv:gr-qc/2006.12577]}}.
\newblock
  doi:{\changeurlcolor{black}\href{https://doi.org/10.3390/sym12081264}{\detokenize{10.3390/sym12081264}}}.

\bibitem[Bardeen(1968)]{Bardeen1968}
Bardeen, J.
\newblock {\em Non-singular general-relativistic gravitational collapse};
  presented at GR5, Tiflis, U.S.S.R., and published in the conference
  proceedings in the U.S.S.R.,  1968.

\bibitem[Dymnikova(1992)]{Dymnikova:1992ux}
Dymnikova, I.
\newblock {Vacuum nonsingular black hole}.
\newblock {\em Gen. Rel. Grav.} {\bf 1992}, {\em 24},~235--242.
\newblock
  doi:{\changeurlcolor{black}\href{https://doi.org/10.1007/BF00760226}{\detokenize{10.1007/BF00760226}}}.

\bibitem[Hayward(2006)]{Hayward2005}
Hayward, S.A.
\newblock {Formation and evaporation of regular black holes}.
\newblock {\em Phys.\ Rev.\ Lett.} {\bf 2006}, {\em 96},~031103.
\newblock
  doi:{\changeurlcolor{black}\href{https://doi.org/10.1103/PhysRevLett.96.031103}{\detokenize{10.1103/PhysRevLett.96.031103}}}.

\bibitem[Bambi and Modesto(2013)]{Bambi:2013ufa}
Bambi, C.; Modesto, L.
\newblock {Rotating regular black holes}.
\newblock {\em Phys. Lett. B} {\bf 2013}, {\em 721},~329--334,
  \href{http://xxx.lanl.gov/abs/1302.6075}{{\normalfont
  [arXiv:gr-qc/1302.6075]}}.
\newblock
  doi:{\changeurlcolor{black}\href{https://doi.org/10.1016/j.physletb.2013.03.025}{\detokenize{10.1016/j.physletb.2013.03.025}}}.

\bibitem[Frolov(2016)]{Frolov:2016pav}
Frolov, V.P.
\newblock {Notes on nonsingular models of black holes}.
\newblock {\em Phys. Rev. D} {\bf 2016}, {\em 94},~104056,
  \href{http://xxx.lanl.gov/abs/1609.01758}{{\normalfont
  [arXiv:gr-qc/1609.01758]}}.
\newblock
  doi:{\changeurlcolor{black}\href{https://doi.org/10.1103/PhysRevD.94.104056}{\detokenize{10.1103/PhysRevD.94.104056}}}.

\bibitem[Abbott(2016)]{Abbott2016}
Abbott, B.P.e.a.
\newblock {Observation of Gravitational Waves from a Binary Black Hole Merger}.
\newblock {\em Phys.\ Rev.\ Lett.} {\bf 2016}, {\em 116},~061102.
\newblock
  doi:{\changeurlcolor{black}\href{https://doi.org/10.1103/PhysRevLett.116.061102}{\detokenize{10.1103/PhysRevLett.116.061102}}}.

\bibitem[Collaboration(2019{\natexlab{a}})]{EHT2019}
Collaboration, E.H.T.
\newblock {First M87 Event Horizon Telescope Results. I. The Shadow of the
  Supermassive Black Hole}.
\newblock {\em Astrophys.\ J.} {\bf 2019}, {\em 875},~L1.
\newblock
  doi:{\changeurlcolor{black}\href{https://doi.org/10.3847/2041-8213/ab0ec7}{\detokenize{10.3847/2041-8213/ab0ec7}}}.

\bibitem[Collaboration(2019{\natexlab{b}})]{paper6}
Collaboration, E.H.T.
\newblock {First M87 Event Horizon Telescope Results. VI. The Shadow and Mass
  of the Central Black Hole}.
\newblock {\em Astrophys. J.} {\bf 2019}, {\em 875},~L6.
\newblock
  doi:{\changeurlcolor{black}\href{https://doi.org/10.3847/2041-8213/ab1141}{\detokenize{10.3847/2041-8213/ab1141}}}.

\bibitem[Giddings(2019)]{Giddings:2019jwy}
Giddings, S.B.
\newblock {Searching for quantum black hole structure with the Event Horizon
  Telescope}.
\newblock {\em Universe} {\bf 2019}, {\em 5},~201,
  \href{http://xxx.lanl.gov/abs/1904.05287}{{\normalfont
  [arXiv:gr-qc/1904.05287]}}.
\newblock
  doi:{\changeurlcolor{black}\href{https://doi.org/10.3390/universe5090201}{\detokenize{10.3390/universe5090201}}}.

\bibitem[Carballo-Rubio \em{et~al.}(2020)Carballo-Rubio, Di~Filippo, Liberati,
  and Visser]{Carballo-Rubio:2019nel}
Carballo-Rubio, R.; Di~Filippo, F.; Liberati, S.; Visser, M.
\newblock {Opening the Pandora\textquoteright{}s box at the core of black
  holes}.
\newblock {\em Class. Quant. Grav.} {\bf 2020}, {\em 37},~145005,
  \href{http://xxx.lanl.gov/abs/1908.03261}{{\normalfont
  [arXiv:gr-qc/1908.03261]}}.
\newblock
  doi:{\changeurlcolor{black}\href{https://doi.org/10.1088/1361-6382/ab8141}{\detokenize{10.1088/1361-6382/ab8141}}}.

\bibitem[Bonanno \em{et~al.}(2020)Bonanno, Khosravi, and
  Saueressig]{Bonanno:2020fgp}
Bonanno, A.; Khosravi, A.P.; Saueressig, F.
\newblock {Regular black holes have stable cores} {\bf 2020}.
\newblock  \href{http://xxx.lanl.gov/abs/2010.04226}{{\normalfont
  [arXiv:gr-qc/2010.04226]}}.

\bibitem[Nicolini \em{et~al.}(2006)Nicolini, Smailagic, and
  Spallucci]{Nicolini:2005vd}
Nicolini, P.; Smailagic, A.; Spallucci, E.
\newblock {Noncommutative geometry inspired Schwarzschild black hole}.
\newblock {\em Phys. Lett.} {\bf 2006}, {\em B632},~547--551,
  \href{http://xxx.lanl.gov/abs/gr-qc/0510112}{{\normalfont
  [arXiv:gr-qc/gr-qc/0510112]}}.
\newblock
  doi:{\changeurlcolor{black}\href{https://doi.org/10.1016/j.physletb.2005.11.004}{\detokenize{10.1016/j.physletb.2005.11.004}}}.

\bibitem[Rovelli and Vidotto(2014)]{Rovelli:2014cta}
Rovelli, C.; Vidotto, F.
\newblock {Planck stars}.
\newblock {\em Int. J. Mod. Phys.} {\bf 2014}, {\em D23},~1442026,
  \href{http://xxx.lanl.gov/abs/1401.6562}{{\normalfont
  [arXiv:gr-qc/1401.6562]}}.
\newblock
  doi:{\changeurlcolor{black}\href{https://doi.org/10.1142/S0218271814420267}{\detokenize{10.1142/S0218271814420267}}}.

\bibitem[Saueressig \em{et~al.}(2016)Saueressig, Alkofer, D'Odorico, and
  Vidotto]{Saueressig2015}
Saueressig, F.; Alkofer, N.; D'Odorico, G.; Vidotto, F.
\newblock {Black holes in Asymptotically Safe Gravity}.
\newblock {\em PoS} {\bf 2016}, {\em FFP14},~174,
  \href{http://xxx.lanl.gov/abs/1503.06472}{{\normalfont
  [arXiv:hep-th/1503.06472]}}.
\newblock
  doi:{\changeurlcolor{black}\href{https://doi.org/10.22323/1.224.0174}{\detokenize{10.22323/1.224.0174}}}.

\bibitem[Litim and Nikolakopoulos(2014)]{Litim:2013gga}
Litim, D.F.; Nikolakopoulos, K.
\newblock {Quantum gravity effects in Myers-Perry space-times}.
\newblock {\em JHEP} {\bf 2014}, {\em 04},~021,
  \href{http://xxx.lanl.gov/abs/1308.5630}{{\normalfont
  [arXiv:hep-th/1308.5630]}}.
\newblock
  doi:{\changeurlcolor{black}\href{https://doi.org/10.1007/JHEP04(2014)021}{\detokenize{10.1007/JHEP04(2014)021}}}.

\bibitem[Pawlowski and Stock(2018)]{Pawlowski:2018swz}
Pawlowski, J.M.; Stock, D.
\newblock {Quantum-improved Schwarzschild-(A)dS and Kerr-(A)dS spacetimes}.
\newblock {\em Phys. Rev.} {\bf 2018}, {\em D98},~106008,
  \href{http://xxx.lanl.gov/abs/1807.10512}{{\normalfont
  [arXiv:hep-th/1807.10512]}}.
\newblock
  doi:{\changeurlcolor{black}\href{https://doi.org/10.1103/PhysRevD.98.106008}{\detokenize{10.1103/PhysRevD.98.106008}}}.

\bibitem[Dymnikova(1992)]{Dymnikova1992}
Dymnikova, I.
\newblock {Vacuum nonsingular black hole}.
\newblock {\em Gen. Rel. Grav.} {\bf 1992}, {\em 24},~235--242.
\newblock
  doi:{\changeurlcolor{black}\href{https://doi.org/10.1007/BF00760226}{\detokenize{10.1007/BF00760226}}}.

\bibitem[Bambi and Modesto(2013)]{Bambi2013}
Bambi, C.; Modesto, L.
\newblock {Rotating regular black holes}.
\newblock {\em Phys.\ Lett.\ B} {\bf 2013}, {\em 721},~329--334.
\newblock
  doi:{\changeurlcolor{black}\href{https://doi.org/10.1016/j.physletb.2013.03.025}{\detokenize{10.1016/j.physletb.2013.03.025}}}.

\bibitem[Newman and Janis(1965)]{Newman1965}
Newman, E.T.; Janis, A.I.
\newblock {Note on the Kerr spinning particle metric}.
\newblock {\em J. Math. Phys.} {\bf 1965}, {\em 6},~915--917.
\newblock
  doi:{\changeurlcolor{black}\href{https://doi.org/10.1063/1.1704350}{\detokenize{10.1063/1.1704350}}}.

\bibitem[Drake and Szekeres(2000)]{Drake:1998gf}
Drake, S.; Szekeres, P.
\newblock {Uniqueness of the Newman-Janis algorithm in generating the
  Kerr-Newman metric}.
\newblock {\em Gen. Rel. Grav.} {\bf 2000}, {\em 32},~445--458,
  \href{http://xxx.lanl.gov/abs/gr-qc/9807001}{{\normalfont [gr-qc/9807001]}}.
\newblock
  doi:{\changeurlcolor{black}\href{https://doi.org/10.1023/A:1001920232180}{\detokenize{10.1023/A:1001920232180}}}.

\bibitem[Goroff and Sagnotti(1985)]{Goroff:1985sz}
Goroff, M.H.; Sagnotti, A.
\newblock {Quantum Gravity at two Loops}.
\newblock {\em Phys. Lett.} {\bf 1985}, {\em 160B},~81--86.
\newblock
  doi:{\changeurlcolor{black}\href{https://doi.org/10.1016/0370-2693(85)91470-4}{\detokenize{10.1016/0370-2693(85)91470-4}}}.

\bibitem[van~de Ven(1992)]{vandeVen:1991gw}
van~de Ven, A.E.M.
\newblock {Two loop quantum gravity}.
\newblock {\em Nucl. Phys.} {\bf 1992}, {\em B378},~309--366.
\newblock
  doi:{\changeurlcolor{black}\href{https://doi.org/10.1016/0550-3213(92)90011-Y}{\detokenize{10.1016/0550-3213(92)90011-Y}}}.

\bibitem[Gies \em{et~al.}(2016)Gies, Knorr, Lippoldt, and
  Saueressig]{Gies:2016con}
Gies, H.; Knorr, B.; Lippoldt, S.; Saueressig, F.
\newblock {Gravitational Two-Loop Counterterm Is Asymptotically Safe}.
\newblock {\em Phys. Rev. Lett.} {\bf 2016}, {\em 116},~211302,
  \href{http://xxx.lanl.gov/abs/1601.01800}{{\normalfont
  [arXiv:hep-th/1601.01800]}}.
\newblock
  doi:{\changeurlcolor{black}\href{https://doi.org/10.1103/PhysRevLett.116.211302}{\detokenize{10.1103/PhysRevLett.116.211302}}}.

\bibitem[Kluth and Litim(2020)]{Kluth:2020bdv}
Kluth, Y.; Litim, D.
\newblock {Fixed Points of Quantum Gravity and the Dimensionality of the UV
  Critical Surface} {\bf 2020}.
\newblock  \href{http://xxx.lanl.gov/abs/2008.09181}{{\normalfont
  [arXiv:hep-th/2008.09181]}}.

\bibitem[Draper \em{et~al.}(2020)Draper, Knorr, Ripken, and
  Saueressig]{Draper:2020bop}
Draper, T.; Knorr, B.; Ripken, C.; Saueressig, F.
\newblock {Finite Quantum Gravity Amplitudes: No Strings Attached}.
\newblock {\em Phys. Rev. Lett.} {\bf 2020}, {\em 125},~181301,
  \href{http://xxx.lanl.gov/abs/2007.00733}{{\normalfont
  [arXiv:hep-th/2007.00733]}}.
\newblock
  doi:{\changeurlcolor{black}\href{https://doi.org/10.1103/PhysRevLett.125.181301}{\detokenize{10.1103/PhysRevLett.125.181301}}}.

\bibitem[Platania and Wetterich(2020)]{Platania:2020knd}
Platania, A.; Wetterich, C.
\newblock {Non-perturbative unitarity and fictitious ghosts in quantum
  gravity}.
\newblock {\em Phys. Lett. B} {\bf 2020}, {\em 811},~135911,
  \href{http://xxx.lanl.gov/abs/2009.06637}{{\normalfont
  [arXiv:hep-th/2009.06637]}}.
\newblock
  doi:{\changeurlcolor{black}\href{https://doi.org/10.1016/j.physletb.2020.135911}{\detokenize{10.1016/j.physletb.2020.135911}}}.

\bibitem[Kamionkowski and March-Russell(1992)]{Kamionkowski:1992mf}
Kamionkowski, M.; March-Russell, J.
\newblock {Planck scale physics and the Peccei-Quinn mechanism}.
\newblock {\em Phys. Lett. B} {\bf 1992}, {\em 282},~137--141,
  \href{http://xxx.lanl.gov/abs/hep-th/9202003}{{\normalfont
  [hep-th/9202003]}}.
\newblock
  doi:{\changeurlcolor{black}\href{https://doi.org/10.1016/0370-2693(92)90492-M}{\detokenize{10.1016/0370-2693(92)90492-M}}}.

\bibitem[Kallosh \em{et~al.}(1995)Kallosh, Linde, Linde, and
  Susskind]{Kallosh:1995hi}
Kallosh, R.; Linde, A.D.; Linde, D.A.; Susskind, L.
\newblock {Gravity and global symmetries}.
\newblock {\em Phys. Rev. D} {\bf 1995}, {\em 52},~912--935,
  \href{http://xxx.lanl.gov/abs/hep-th/9502069}{{\normalfont
  [hep-th/9502069]}}.
\newblock
  doi:{\changeurlcolor{black}\href{https://doi.org/10.1103/PhysRevD.52.912}{\detokenize{10.1103/PhysRevD.52.912}}}.

\bibitem[Banks and Seiberg(2011)]{Banks:2010zn}
Banks, T.; Seiberg, N.
\newblock {Symmetries and Strings in Field Theory and Gravity}.
\newblock {\em Phys. Rev. D} {\bf 2011}, {\em 83},~084019,
  \href{http://xxx.lanl.gov/abs/1011.5120}{{\normalfont
  [arXiv:hep-th/1011.5120]}}.
\newblock
  doi:{\changeurlcolor{black}\href{https://doi.org/10.1103/PhysRevD.83.084019}{\detokenize{10.1103/PhysRevD.83.084019}}}.

\bibitem[Eichhorn and Held(2017)]{Eichhorn:2017eht}
Eichhorn, A.; Held, A.
\newblock {Viability of quantum-gravity induced ultraviolet completions for
  matter}.
\newblock {\em Phys. Rev. D} {\bf 2017}, {\em 96},~086025,
  \href{http://xxx.lanl.gov/abs/1705.02342}{{\normalfont
  [arXiv:gr-qc/1705.02342]}}.
\newblock
  doi:{\changeurlcolor{black}\href{https://doi.org/10.1103/PhysRevD.96.086025}{\detokenize{10.1103/PhysRevD.96.086025}}}.

\bibitem[Eichhorn and Pauly(2020)]{Eichhorn:2020sbo}
Eichhorn, A.; Pauly, M.
\newblock {Constraining power of asymptotic safety for scalar fields} {\bf
  2020}.
\newblock  \href{http://xxx.lanl.gov/abs/2009.13543}{{\normalfont
  [arXiv:hep-th/2009.13543]}}.

\bibitem[Feldbrugge \em{et~al.}(2017)Feldbrugge, Lehners, and
  Turok]{Feldbrugge:2017kzv}
Feldbrugge, J.; Lehners, J.L.; Turok, N.
\newblock {Lorentzian Quantum Cosmology}.
\newblock {\em Phys. Rev. D} {\bf 2017}, {\em 95},~103508,
  \href{http://xxx.lanl.gov/abs/1703.02076}{{\normalfont
  [arXiv:hep-th/1703.02076]}}.
\newblock
  doi:{\changeurlcolor{black}\href{https://doi.org/10.1103/PhysRevD.95.103508}{\detokenize{10.1103/PhysRevD.95.103508}}}.

\bibitem[Horava(2009)]{Horava2009}
Horava, P.
\newblock {Quantum Gravity at a Lifshitz Point}.
\newblock {\em Phys. Rev. D} {\bf 2009}, {\em 79},~084008.
\newblock
  doi:{\changeurlcolor{black}\href{https://doi.org/10.1103/PhysRevD.79.084008}{\detokenize{10.1103/PhysRevD.79.084008}}}.

\bibitem[Steinwachs(2020)]{Steinwachs:2020jkj}
Steinwachs, C.F.
\newblock {Towards a unitary, renormalizable and ultraviolet-complete quantum
  theory of gravity} {\bf 2020}.
\newblock  \href{http://xxx.lanl.gov/abs/2004.07842}{{\normalfont
  [arXiv:hep-th/2004.07842]}}.
\newblock
  doi:{\changeurlcolor{black}\href{https://doi.org/10.3389/fphy.2020.00185}{\detokenize{10.3389/fphy.2020.00185}}}.

\bibitem[Daum and Reuter(2013)]{Daum:2013fu}
Daum, J.; Reuter, M.
\newblock {Einstein-Cartan gravity, Asymptotic Safety, and the running Immirzi
  parameter}.
\newblock {\em Annals Phys.} {\bf 2013}, {\em 334},~351--419,
  \href{http://xxx.lanl.gov/abs/1301.5135}{{\normalfont
  [arXiv:hep-th/1301.5135]}}.
\newblock
  doi:{\changeurlcolor{black}\href{https://doi.org/10.1016/j.aop.2013.04.002}{\detokenize{10.1016/j.aop.2013.04.002}}}.

\bibitem[Harst and Reuter(2012)]{Harst:2012ni}
Harst, U.; Reuter, M.
\newblock {The 'Tetrad only' theory space: Nonperturbative renormalization flow
  and Asymptotic Safety}.
\newblock {\em JHEP} {\bf 2012}, {\em 05},~005,
  \href{http://xxx.lanl.gov/abs/1203.2158}{{\normalfont
  [arXiv:hep-th/1203.2158]}}.
\newblock
  doi:{\changeurlcolor{black}\href{https://doi.org/10.1007/JHEP05(2012)005}{\detokenize{10.1007/JHEP05(2012)005}}}.

\bibitem[Knorr and Ripken(2020)]{Knorr:2020bjm}
Knorr, B.; Ripken, C.
\newblock {Scattering amplitudes in affine gravity} {\bf 2020}.
\newblock  \href{http://xxx.lanl.gov/abs/2012.05144}{{\normalfont
  [arXiv:hep-th/2012.05144]}}.

\bibitem[de~Brito and Pereira(2020)]{deBrito:2020rwu}
de~Brito, G.P.; Pereira, A.D.
\newblock {Unimodular quantum gravity: Steps beyond perturbation theory}.
\newblock {\em JHEP} {\bf 2020}, {\em 09},~196,
  \href{http://xxx.lanl.gov/abs/2007.05589}{{\normalfont
  [arXiv:hep-th/2007.05589]}}.
\newblock
  doi:{\changeurlcolor{black}\href{https://doi.org/10.1007/JHEP09(2020)196}{\detokenize{10.1007/JHEP09(2020)196}}}.

\end{thebibliography}

\end{document}